%% file: NPS-25-002_temp.tex
\documentclass[11pt,twoside,a4paper,cmspaper,final,collab]{cms-tdr}
\def\svnVersion{6014af4}\def\svnDate{2026/10/31}\def\cmsCernNoTag{CERN-EP-2026-098}\def\cmsCernDate{\today}\def\cmsMessage{Published in the Journal of High Energy Physics as \href{http://dx.doi.org/10.1007/JHEP10(2026)249}{\doi{10.1007/JHEP10(2026)249}.}}
\begin{document}\cmsNoteHeader{NPS-25-002}

\newcommand{\sqs}[1]{\ensuremath{\sqrt{s}=#1\TeV}\xspace}
\providecommand{\PA}{{\HepParticle{A}{}{}}\xspace}
\providecommand{\PSHmp}{{\HepParticle{\PSH}{}{\mp}}\xspace}
\providecommand{\PGgst}{\HepParticle{\PGg}{}{\ast}\xspace}
\newcommand{\mch}{\ensuremath{m_{\PSHpm}}\xspace}
\newcommand{\ma}{\ensuremath{m_{\PA}}\xspace}
\newcommand{\mh}{\ensuremath{m_{\PH}}\xspace}
\newcommand{\mdiff}{\ensuremath{\ma -\mh}\xspace}
\newcommand{\mll}{\ensuremath{m_{\Pell\Pell}}\xspace}
\newcommand{\ee}{\ensuremath{\Pep\Pem}\xspace}
\newcommand{\mumu}{\ensuremath{\PGmp\PGmm}\xspace}
\newcommand{\pnn}{\ensuremath{\mathrm{pNN}(\mh,\ma)}\xspace}
\newcommand{\invlumiII}{138}
\newcommand{\invlumiIII}{35}
\newcommand{\Zjets}{\ensuremath{\PZ\text{+jets}}\xspace}
\newcommand{\phojets}{\ensuremath{\gamma\text{+jets}}\xspace}
\newcommand{\DYjets}{\ensuremath{\PZ/\PGgst\text{+jets}}\xspace}
\newcommand{\Wlvjets}{\ensuremath{\PW(\Pell\PGn)\text{+jets}}\xspace}
\newcommand{\NNPDF}{\textsc{nnpdf}\xspace}
\newcommand{\pp}{\ensuremath{\Pp\Pp}\xspace}

\cmsNoteHeader{NPS-25-002}
\title{Search for pair production of additional neutral scalars within the Inert Doublet Model in a final state with two electrons or two muons in proton-proton collisions at \texorpdfstring{\sqs{13}}{sqrt(s) = 13 TeV} and \texorpdfstring{13.6\TeV}{13.6 TeV}}

\date{\today}

\abstract{A first dedicated search for pair production of new scalars predicted by the Inert Doublet Model is performed using proton-proton collisions. Data were collected with the CMS detector at the CERN LHC at \sqs{13} and 13.6\TeV, corresponding to integrated luminosities of \invlumiII\fbinv and \invlumiIII\fbinv, respectively. Within this model, four additional scalar bosons (\PH, \PA, \PSHp, and  \PSHm) are predicted. Through an additional discrete symmetry, the lightest new scalar, \PH, is stable, rendering it a viable dark matter candidate. These candidates can originate from quark-antiquark annihilation producing an offshell \PZ boson that decays to a pair of the new scalars. The target final state consists of exactly two opposite-charge same-flavour leptons (electrons or muons), with missing transverse momentum due to the stable neutral scalars, and very little hadronic activity. A parameterised neural network is used to separate the signal from the standard model background. No significant excess of events is observed. Exclusion limits at 95\% confidence level are set on the production cross section of the two new neutral scalars, \PH and \PA, expressed in terms of their masses, \mh and \ma, in the \mh \vs \mdiff plane. The observed (expected) exclusion region reaches $\mh=108\,(106)\GeV$ for $\mdiff=78\,(76)\GeV$ and at $\mh=70\GeV$, covers the range of $\mdiff=40$--90\,(35--90)\GeV.}

\hypersetup{
  pdfauthor={CMS Collaboration},
  pdftitle={Search for pair production of additional neutral scalars within the Inert Doublet Model in a final state with two electrons or two muons in proton-proton collisions at sqrt(s) = 13 TeV and 13.6 TeV},
  pdfsubject={CMS},
  pdfkeywords={CMS, BSM, Inert Doublet Model}
}

\maketitle 

\section{Introduction}

The standard model (SM) of particle physics is the most successful
framework for describing the behaviour of the known
particles. Nevertheless, the SM is unable to account for numerous
observations. In particular, multiple cosmological measurements point
towards the existence of dark matter (DM)
\cite{ParticleDataGroup:2024cfk}, the abundance of which is predicted
to significantly outweigh that of normal matter. Although the specific
nature of DM remains unknown, one of the most widely supported
scenarios is that DM takes the form of a weakly interacting,
long-lived massive
particle~\cite{ParticleDataGroup:2024cfk,Roszkowski_2018}. The SM,
however, contains no such particle.  Assuming that DM is a particle
that can interact with normal matter, it might be possible to detect
it experimentally. This can happen in a number of
ways~\cite{ParticleDataGroup:2024cfk}. One possibility is that stable
DM particles passing through our solar system could directly interact
with normal matter, imparting energy that can be detected in
direct-detection experiments. Another possibility is through the
detection of SM particles produced through annihilation of DM
particles. Alternatively, DM could be produced at a collider, leading,
\eg to a missing energy
signature~\cite{ATLAS:2024fdw,CMS:2024zqs}. This paper focuses on the
latter signature, searching for DM production at the CERN LHC.

The search for a massive DM candidate has naturally led to the development of
several beyond-the-SM theories. One particular model is the Inert Doublet
Model (IDM), a two-Higgs-doublet model with an additional, unbroken
\(\mathbb{Z}_2\) symmetry
\cite{Deshpande:1977rw,LopezHonorez:2006gr,Barbieri:2006dq,Cao:2007rm,Ilnicka:2015jba,Kalinowski:2020rmb}.
As a result of the \(\mathbb{Z}_2\) symmetry, additional particles from the
second Higgs doublet do not couple directly to the fermionic fields, whilst
also rendering the lightest of these new particles stable. Ultimately, this
introduces a weakly interacting, massive, stable particle, which is a viable
DM candidate. Although many phenomenological studies exist that
explore the sensitivity at future colliders
\cite{LopezHonorez:2006gr,Kalinowski:2018kdn,Zarnecki:2020swm,Braathen:2024lyl,Klamka:2022ukx,Belyaev:2021ngh,Kalinowski:2020rmb,Guo-He:2020nok,Kalinowski:2018ylg,Datta:2016nfz,Ilnicka:2015jba,Hashemi:2015swh,Abouabid:2020eik,Abouabid:2022rnd,Bal:2025nbu,Ghosh:2025dcv},
no dedicated search for IDM scalars has yet been performed with experimental
data. Existing experimental analyses with collider data are generally not
sensitive to a large part of the parameter space, for example because of high threshold
requirements on the missing transverse momentum (\ptmiss)~\cite{Sekmen:2025bxv}, and
their reinterpretation under the IDM hypothesis leads to very loose constraints
\cite{Dercks:2018wch,Belyaev:2022wrn,PhysRevD.91.115011,Lahiri:2025opz}.

In this paper, we present a search for IDM signatures in proton-proton
(\pp) collisions at centre-of-mass energies of 13\TeV and 13.6\TeV
delivered by the LHC between 2016 and 2022, and recorded by the CMS
experiment. This corresponds to total integrated luminosities of
\invlumiII\fbinv and \invlumiIII\fbinv at \sqs{13} and 13.6\TeV,
respectively. The main target process is the production $\Pp \Pp
\to \PA \PH$, where $\PA$ and $\PH$ are the additional neutral
IDM scalars of which $\PH$ is the dark matter candidate. The scalar
$\PA$ subsequently decays into $\PZ\PH$, and the final state studied
contains exactly one opposite-charge same-flavour (OCSF) lepton pair
(electrons or muons) accompanied by \ptmiss from
the two escaping $\PH$ particles. For an invariant mass of the
dilepton system $\mll$ below 12\GeV and above 80\GeV we have limited 
sensitivity due to the low-mass resonances (\PJGy, \PGU, etc.)  and the large resonant production of the Z boson, respectively. 
For this reason, this analysis focuses on
events with $12<\mll<80\GeV$. The search
is performed over the ranges $\mh \in [60, 180]\GeV$ and $\mdiff \in
[20, 100]\GeV$, chosen to cover the region of parameter space not
excluded by previous theoretical and experimental constraints detailed
in Section~\ref{sec:idm}.

This paper is organised as follows. The IDM is introduced in
Section~\ref{sec:idm}, as well as theoretical assumptions that
reduce the number of free parameters in the interpretation.
Section~\ref{sec:CMSdet} introduces the CMS detector. The data and the
simulated samples for the signal and backgrounds are detailed in
Section~\ref{sec:mc}, with the object definitions given in
Section~\ref{sec:objects}. A set of selections that remove the majority of the
SM backgrounds is defined in Section~\ref{sec:sel}, followed by a description
of a multivariate analysis using a parameterised neural network to further
discriminate between signal and background. The main backgrounds are estimated
using data in dedicated control regions (CRs) as described in Section~\ref{sec:bkg}.
Section~\ref{sec:systs} describes the systematic uncertainties.
Finally, the results are presented in Section~\ref{sec:res} and the paper is summarised in 
Section~\ref{sec:concl}. Tabulated results are accessible from the HEPData record for
this analysis~\cite{hepdata}.

\section{The Inert Doublet Model}
\label{sec:idm}

The Inert Doublet Model is a two-Higgs-doublet model with an additional,
unbroken $\mathbb{Z}_2$
symmetry~\cite{Deshpande:1977rw,LopezHonorez:2006gr,Barbieri:2006dq,Cao:2007rm,Ilnicka:2015jba,Kalinowski:2020rmb}.
Following the convention detailed in Ref.~\cite{Ilnicka:2015jba}, the
doublets, denoted $\phi_{\mathrm{D}}$ and $\phi_{\mathrm{S}}$, have the following transformation
properties under the $\mathbb{Z}_2$ symmetry:
\begin{equation}
  \mathbb{Z}_2 : \, \phi_{\mathrm{D}} \longrightarrow -\phi_{\mathrm{D}}, \;
  \phi_{\mathrm{S}} \longrightarrow \phi_{\mathrm{S}}, \; SM \longrightarrow SM,
\end{equation}
where $SM$ stands for all other SM fields.

Because of the $\mathbb{Z}_2$ symmetry remaining unbroken during the
spontaneous symmetry breaking mechanism, $\phi_{\mathrm{D}}$ is forced
to have zero vacuum expectation value, whereas $\phi_{\mathrm{S}}$
undergoes the same mechanism as in the SM, leading to the usual
125\GeV SM Higgs boson, $\Ph$. This configuration defines the inert
vacuum~\cite{Ginzburg:2010wa}. The $\phi_{\mathrm{D}}$ doublet leads
to four new scalars $\PH, \PA,  \PSHpm$, with masses \mh, \ma, and \mch,
respectively. In addition to the masses of the IDM particles, two
coupling constants are chosen, leading to the following five free
parameters in the IDM: $\mh$, $\ma$, $\mch$, $\lambda_{2}$, and
$\lambda_{345}$.  Here, $\lambda_{2}$ is the coupling constant for the
quartic vertex between four IDM particles, and $\lambda_{345}$ is the
coupling constant for the $\Ph\PH\PH$ vertex. The couplings between
the new scalars and the SM gauge bosons are determined solely by
quantities from the SM electroweak (EW) sector. The exact IDM vertex
factors are presented in \eg Ref.~\cite{Dercks:2018wch}. The scalars from
the two doublets do not mix because of the $\mathbb{Z}_2$ symmetry.

The $\mathbb{Z}_2$ symmetry forbids any direct coupling between the SM fermion
fields and the IDM particles, and as such the IDM particles only couple
directly to bosons. Furthermore, as the symmetry is exact, remaining unbroken
after the symmetry breaking mechanism, IDM particles are always produced in
pairs. Consequently, the lightest of the IDM particles is stable, thereby
providing a viable DM candidate. Following the conventions in
Refs.~\cite{Kalinowski:2020rmb,Ilnicka:2015jba}, $\PH$ is chosen to be the
lightest among $\PH$, $\PA$, and $ \PSHpm$.
Unlike other two-Higgs-doublet models, $\PH/\PA$ do not correspond to
scalar/pseudoscalar particles as $\phi_{\mathrm{D}}$ has no direct coupling to the
fermions. As a result, the exact roles of $\PH$ and $\PA$ can be exchanged
under a suitable redefinition of the coupling constants, and, therefore, the
choice of which neutral particle is lightest is arbitrary.

\begin{figure}[ht!]
  \centering
  \includegraphics[width=0.3\textwidth]{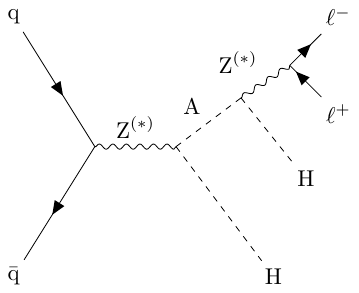}
  \hfill\includegraphics[width=0.3\textwidth]{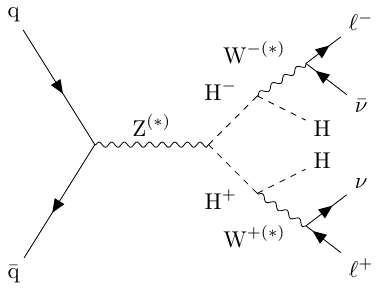}
  \hfill\includegraphics[width=0.3\textwidth]{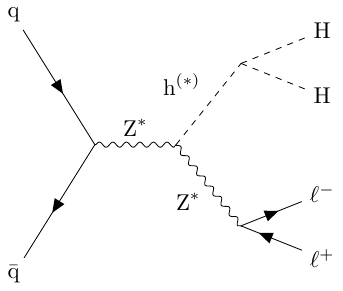}
  \caption{\label{fig:idm_diagram} Leading-order Feynman diagrams
    of: (left) $\PH\PH\Pellp\Pellm$ production through $\PA\PH$ production,
    (middle) $\PH\PH\Pellp\Pellm\PGn\PAGn$ through $ \PSHpm \PSHmp$ production, and (right)
    $\PH\PH\Pellp\Pellm$ through $\Ph\PZ$ production.  }
\end{figure}

At a collider, a typical IDM signature is that of SM EW gauge bosons
with \ptmiss from the DM particles. Possible final
states include monojet, monophoton, invisible Higgs boson decays, or
multilepton channels (see \eg
Refs~\cite{Belyaev:2018ext,Belyaev:2016lok}). This analysis focuses on
the OCSF dilepton (electrons or muons) plus \ptmiss final state, with example production mechanisms shown in
Fig.~\ref{fig:idm_diagram}.

The IDM free parameters are subject to numerous theoretical and
experimental constraints. In this work, the constraints that are
applied follow Ref.~\cite{Ilnicka:2015jba}, with more recent updates
described in
Refs.~\cite{Ilnicka:2018def,Kalinowski:2018ylg,Dercks:2018wch,Kalinowski:2020rmb,Braathen:2024lyl,Robens:2026yar}.
The two-loop constraints on the scalar couplings presented in
Ref.~\cite{Braathen:2024lyl} are not applied, as the code is not
publicly available, instead the leading-order (LO) values for constraints
are used.  Constraints on the potential stemming from vacuum
stability, perturbativity, and perturbative unitarity are included,
together with EW precision observables evaluated using
\textsc{2HDMC}~\cite{Eriksson:2010zzb}. Results for the EW precision
constraints are tested using the oblique parameters
$S,\,T,\,U$~\cite{Altarelli:1990zd,
  Peskin:1990zt,Peskin:1991sw,Maksymyk:1993zm} and compared to the
latest PDG values~\cite{ParticleDataGroup:2024cfk}. All points in the
parameter space are required to reside in the inert
vacuum. Cases with a possible on-shell decay of
the Higgs boson to IDM particles must satisfy the most recent bound
on SM Higgs boson decays to invisible particles, an observed
upper limit at 95\% confidence level on the branching ratio
$\mathcal{B}(\Ph \to \text{inv})<
0.11$~\cite{ATLAS:2023tkt}. Using \textsc{MicrOMEGAs}~\cite{Belanger:2018ccd}, agreement with the upper bound for the relic density as measured by
the Planck experiment is imposed~\cite{Planck:2018vyg}. A detailed
discussion of regions where the IDM can match the observed relic
density can be found in Ref.~\cite{Kalinowski:2020rmb}. Direct-detection
constraints are taken from the LZ
Collaboration~\cite{LZ:2024zvo}. Finally, the mass hierarchy is
required to obey the constraints from searches for supersymmetric
neutralinos at the CERN LEP reinterpreted within the
IDM~\cite{Lundstrom:2008ai} (referred to as LEP reinterpretation in
the following), and that the new particles would not contribute to the
decay widths of the EW gauge bosons at LO.

For the OCSF dilepton final state, the signal cross section is usually
dominated by the $\PA\PH$ production shown in
Fig.~\ref{fig:idm_diagram}, left. As this process has only SM
couplings, the sensitivity predominantly depends on \mh and \mdiff. At
tree level, the quartic coupling between IDM bosons, $\lambda_2$, does
not contribute, and as such the choice of $\lambda_2$ is irrelevant
for signal production. Diagrams involving $\lambda_{345}$, which
couples the IDM particles to the SM Higgs boson as shown in
Fig.~\ref{fig:idm_diagram}, right, are strongly suppressed for
$\mh>m_\Ph/2$, the target region for this analysis.  In addition,
strong upper constraints on the value of $\lambda_{345}$ stem from
direct-detection experiments~\cite{Ilnicka:2015jba,Robens:2026yar},
meaning that this parameter cannot be arbitrarily large. Even
considering the largest value allowed today, the sensitivity of this
analysis is not affected, and $\lambda_{345}$ is set to 10$^{-6}$. The
IDM mass \mch has a lower limit of $\mch = \ma$ and an approximate
upper limit of $\mch = \ma + 50\GeV$ within the range of \mh and \ma
values considered in this analysis. This parameter contributes to the
dilepton signal primarily through the diagram in
Fig.~\ref{fig:idm_diagram}, middle. However, this process generally
has a cross section over an order of magnitude smaller than the
$\PA\PH$ production, as well as poorer discrimination against the SM
background. By varying \mch within the allowed range, no effect is
found on the sensitivity. For the signal production, the choice is
made to fix \mch to $\ma + 50\GeV$. With these assumptions, the number
of free parameters is brought down to two and a scan is performed over
\mh and \mdiff.

\section{The CMS detector}
\label{sec:CMSdet}

The central feature of the CMS apparatus is a superconducting solenoid of
6\unit{m} internal diameter, providing a magnetic field of 3.8\unit{T}. Within
the solenoid volume are a silicon pixel and strip tracker, a lead tungstate
crystal electromagnetic calorimeter (ECAL), and a brass and scintillator hadron
calorimeter, each composed of a barrel and two endcap sections. Forward
calorimeters extend the pseudorapidity $\eta$ coverage provided by the barrel
and endcap detectors. Muons are measured in gas-ionisation detectors embedded
in the steel flux-return yoke outside the solenoid.

Events of interest are selected using a two-tiered trigger
system~\cite{Khachatryan:2016bia,CMS:2024aqx}. The first level (L1) is
composed of custom hardware processors, which use information from the
calorimeters and muon detectors to select events at a rate of about
100\unit{kHz}~\cite{Sirunyan:2020zal}. The second level, known as the
high-level trigger (HLT), is a software-based system that runs a version of the full
event reconstruction optimised for fast processing, reducing the event rate to
about 1 to 5\unit{kHz}.

A detailed description of the CMS detector, together with a definition
of the coordinate system used and the relevant kinematic variables, is
given in Ref.~\cite{Chatrchyan:2008zzk}. The upgraded configuration of
the detector for the LHC Run 3 is given in Ref.~\cite{CMS:2023gfb}.

\section{Data and simulated samples}
\label{sec:mc}

Data were recorded with several triggers, as detailed in Section~\ref{sec:sel},
during the LHC Run 2 (2016--2018) and part of Run 3 (2022), for integrated
luminosities corresponding to \invlumiII\fbinv and \invlumiIII\fbinv, respectively.

\begin{figure}[ht!]
  \centering
  \includegraphics[width=0.49\textwidth]{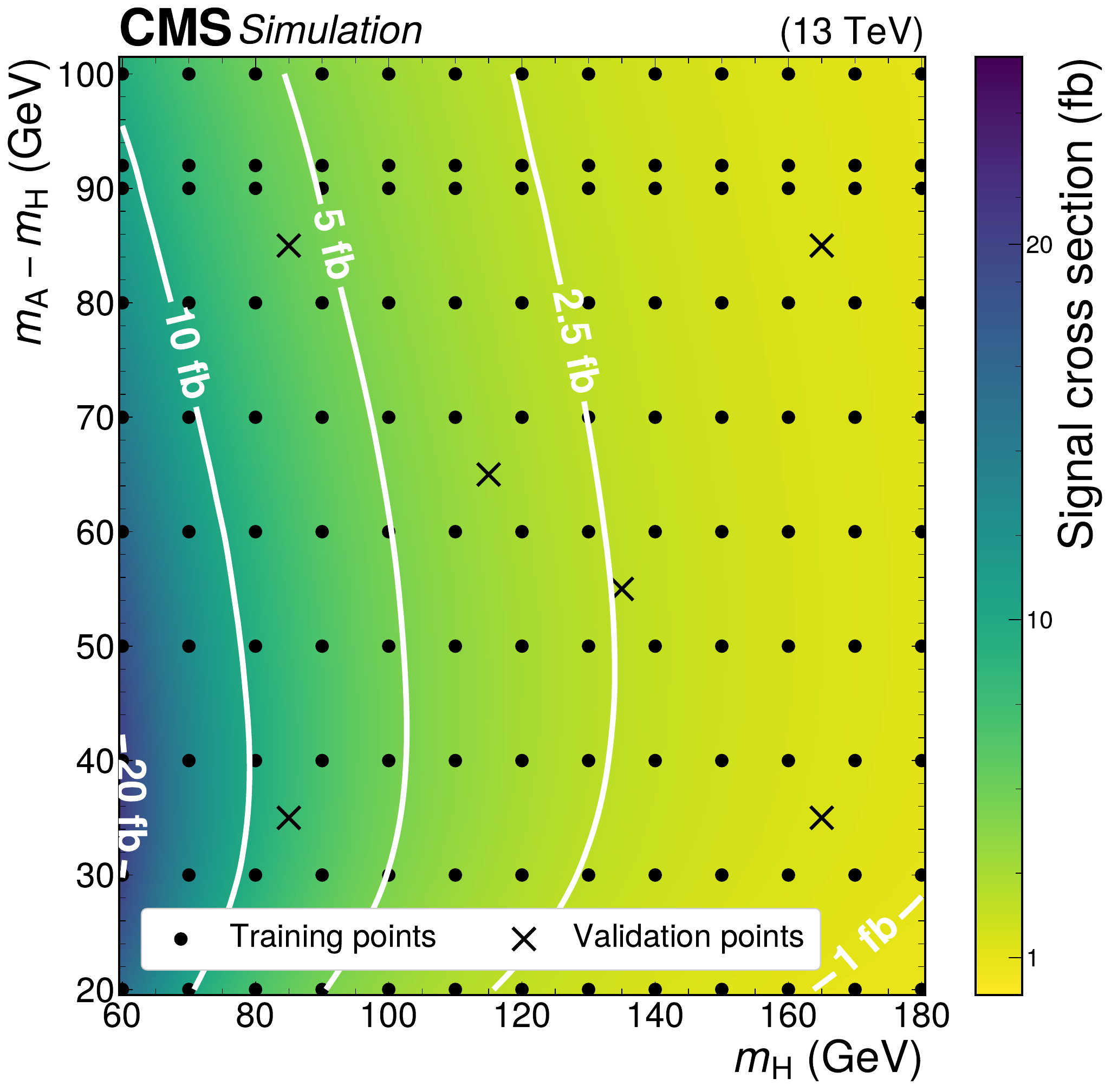}
  \includegraphics[width=0.49\textwidth]{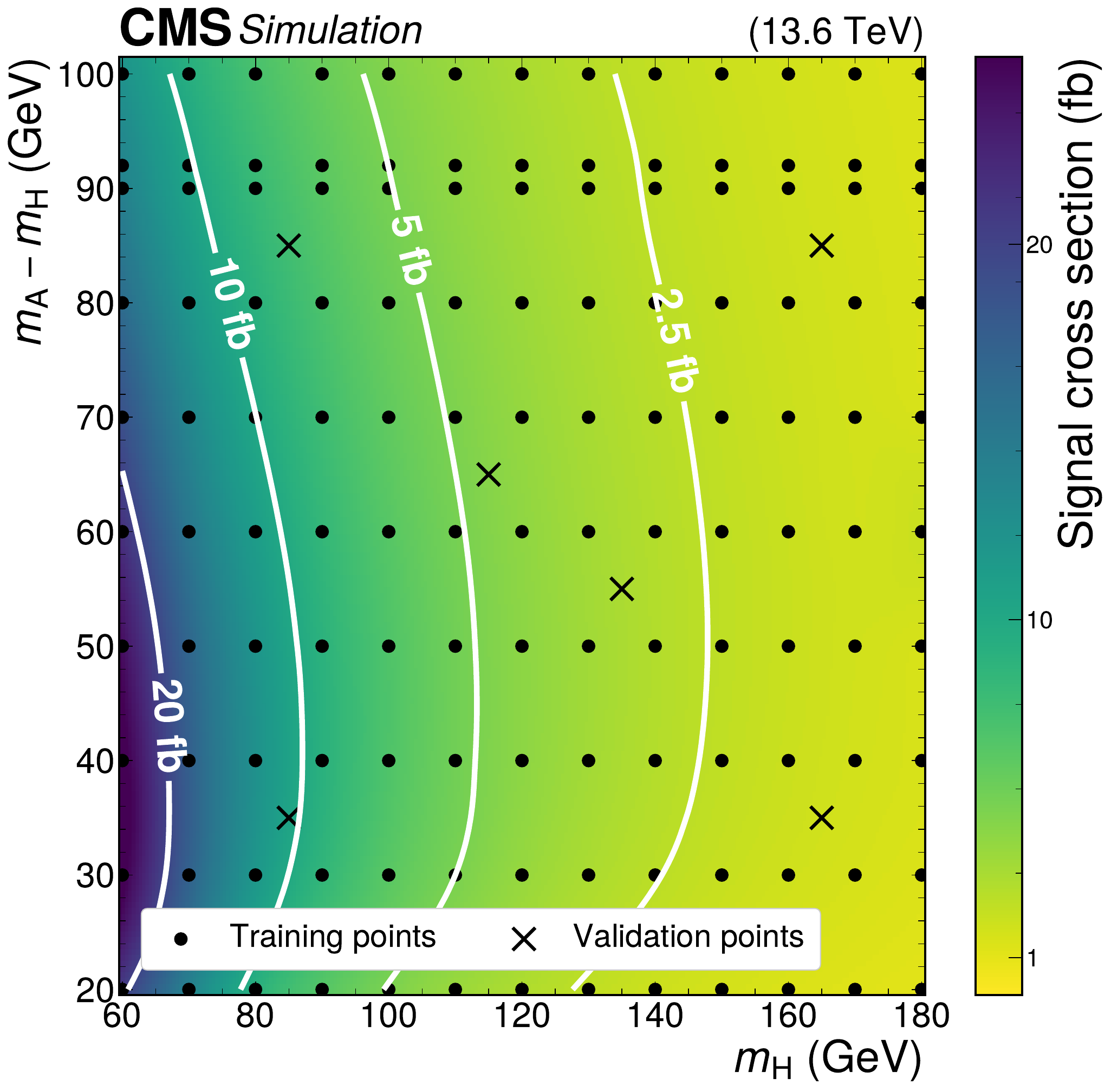}
  \caption{ Total signal cross section at LO obtained with \MGvATNLO
    in the $(\mh, \ma-\mh)$ plane at (left) \sqs{13} and (right)
    \sqs{13.6}, computed for leptons with transverse momentum $\pt >
    10\GeV$. The simulated training points (circles) are used for
    training the classifier and for the final statistical
    inference. The validation points (crosses) are used for testing
    the interpolation of the signal.}
  \label{fig:parameter_scan_mH_mA-mH}
\end{figure}

The IDM signal samples were generated at LO using \MGvATNLO
v2.6.7~\cite{Alwall:2014hca}. The input model is taken from
Ref.~\cite{Goudelis:2013uca}. Instead of targeting individual pair
production modes, two final states are generated directly, such that
all contributing diagrams with the proper interferences are taken into
account. The two final states are $\PH\PH\Pell\Pell$ and
$\PH\PH\Pell\Pell\PGn\PAGn$, where $\Pell=\Pe,\PGm,$ or
$\PGt$. Typically, the former final state is obtained with virtual or
on-shell \PZ bosons, as shown in Fig.~\ref{fig:idm_diagram}, left, and
the sensitivity will depend on the difference in mass between the \PH
and the \PA scalar bosons. The latter subdominant final state is
obtained through a pair of \PW bosons, such as in
Fig.~\ref{fig:idm_diagram}, middle, and the sensitivity will instead
depend on the difference in mass between \PH and  \PSHpm. Samples are
generated over the range $\mh \in [60, 180]$\GeV in steps of 10\GeV,
and $\ma \in [90, 280]$\GeV in steps of 10\GeV, all while maintaining
the inequalities $\mh < \ma$ and $\ma - \mh < 100$\GeV, corresponding
to the region of parameter space where the analysis is expected to be
sensitive.  Furthermore, additional points are simulated at
$\ma=\mh+92$\GeV to account for the kinematic discontinuity at $\mdiff
= m_{\PZ}$ from the $\PZ$ boson being produced on-shell. The signal
cross sections are shown in Fig.~\ref{fig:parameter_scan_mH_mA-mH} in
the $(\mh,\ma-\mh)$ plane for the two $\sqrt{s}$ values, computed for
leptons with transverse momentum $\pt > 10\GeV$ with a numerical
integration error smaller than $0.1\%$. The increase in $\sqrt{s}$
yields a factor of $\approx$1.3 increase in cross section across the
parameter space. The simulated points, shown as the training points in
Fig.~\ref{fig:parameter_scan_mH_mA-mH}, are used for training a
classifier (Section \ref{sec:pnn}) and for the statistical
inference. Additional points, referred to as the validation points in
Fig.~\ref{fig:parameter_scan_mH_mA-mH}, are also simulated to quantify
the interpolation uncertainties (Section \ref{sec:pnn}). These are
produced at points $(\mh, \ma)=\{(85, 120), (85, 170), (115, 180),
(135, 190), (165, 200), (165, 250)\}$\GeV, specifically chosen to
cover a large area of the parameter space.

Several SM processes contribute to producing a pair of OCSF electrons
or muons.  The dominant backgrounds arise from Drell--Yan (DY) production
$\PZ/\PGgst \to \Pell\Pell$ in association with jets,
$\PW$+jets production with a jet being misidentified as a lepton,
diboson production, as well as top quark-antiquark, $\ttbar$, and
single top quark production.

The $\PZ/\PGgst(\Pell\Pell)$+jets and \Wlvjets (with $\Pell = \Pe,
\PGm,$ or $\PGt$) processes are simulated at LO with \MGvATNLO
v2.6.5~\cite{Alwall:2014hca} with up to four additional partons in the
matrix element (ME) calculations, using the MLM jet matching
procedure~\cite{Alwall:2007fs}, with a total cross section scaled to
the next-to-next-to-LO (NNLO) calculations from \FEWZ
v3.1~\cite{Li:2012wna}. For $\PZ/\PGgst(\Pell\Pell)$+jets, events
with a dilepton invariant mass $\mll > 10$\GeV are simulated. Finally,
for $\PZ/\PGgst(\Pell\Pell)$+jets, additional samples are generated
at next-to-LO (NLO) for deriving the NLO/LO multiplicative factors. The NLO samples are
produced with up to two additional partons in the ME calculations using \MGvATNLO v2.6.5 with
the FxFx jet merging scheme~\cite{Alwall:2014hca,Frederix:2012ps}, and
the NLO/LO factors are implemented as a 2-dimensional reweighting as
a function of the generator-level \PZ boson \pt and
generator-level dilepton \mll.

Samples of quark-initiated $\PW\PZ$ events are simulated using
\MGvATNLO v2.6.5~\cite{Alwall:2014hca,Frederix:2012ps} at NLO in perturbative
quantum chromodynamics (pQCD), whereas quark-initiated
$\PW\PW$ and $\PZ\PZ$ processes are simulated at NLO in pQCD using
\POWHEG v2.0~\cite{Nason:2004rx,Frixione:2007vw,Alioli:2010xd,Melia:2011tj}. Diboson production from gluon-gluon fusion ($\Pg
  \Pg \to 2\Pe2\PGn$, $2\PGm2\PGn$, $2\Pe2\PGm$, $2\Pe2\PGt$,
$2\PGm2\PGt$, $4\Pe$, $4\PGm$, $4\PGt$) is simulated with \MCFM v7~\cite{mcfm}.

The \ttbar and single top quark background samples are produced at NLO in pQCD
using \POWHEG v2.0 and v1.0,
respectively~\cite{Campbell:2014kua,Alioli:2009je,Re:2010bp}. The \ttbar cross
section is calculated at NNLO in pQCD and includes resummation of
next-to-next-to-leading logarithmic soft-gluon terms using the \textsc{Top++} v2.0
program~\cite{Top++}. Single top quark processes are also normalised to available
NNLO calculations~\cite{Campbell:2020fhf,Kidonakis:2021vob}.

The following subdominant contributions are also considered: {\ttbar}\PW,
{\ttbar}\PZ and triboson ({\PW}{\PW}\PW, {\PW}{\PW}\PZ, {\PW}{\PZ}\PZ)
production, simulated with \MGvATNLO v2.6.5 at
NLO in pQCD~\cite{Alwall:2014hca,Frederix:2012ps}.

All signal and background processes are simulated using similar Monte
Carlo (MC) generator configurations, utilising the \NNPDF
3.1~\cite{Ball:2014uwa} NNLO parton distribution functions. All
generators are interfaced with \PYTHIA 
v8.240~\cite{Sjostrand:2014zea} (v8.306~\cite{Bierlich:2022pfr}) for
the parton shower simulation, hadronisation, and fragmentation
processes for the Run 2 (Run 3) samples, using the \textsc{CP5} underlying
event tune~\cite{CMS:2019csb}. Interactions of the final-state particles
with the CMS detector are simulated using
\GEANTfour~\cite{Agostinelli:2002hh}. Additional \pp interactions
(pileup) are included in the simulation, and simulated events are
weighted to reproduce the pileup distribution observed in data. The
average number of interactions per crossing is 27 in 2016 data, 38 in
2017, 37 in 2018, and 46 in 2022.

\section{Event reconstruction}
\label{sec:objects}

A global event description is available using the particle-flow (PF)
algorithm~\cite{Sirunyan:2017ulk}. Using a combination of the information
provided by the tracker, calorimeters, and muon systems, the PF algorithm aims
to reconstruct individual particles (PF candidates), classifying them as
electrons, photons, muons, charged hadrons, or neutral hadrons.

The primary vertex is taken to be the vertex corresponding to the hardest
scattering in the event, evaluated using tracking information alone, as
described in Section 9.4.1 of Ref.~\cite{CMS-TDR-15-02}.

Electron candidates~\cite{CMS:2020uim} are selected in the range
$\abs{\eta}<2.5$, ignoring those that fall within the transition
region between the ECAL barrel and endcap components, $1.444 <
\abs{\eta} < 1.566$. Muon candidates~\cite{CMS:2018rym} are selected
with $\abs{\eta}< 2.4$. For electrons (muons), loose and tight working
points are defined in Ref.~\cite{CMS:2020uim}
(Ref.~\cite{CMS:2018rym}) using selection criteria on discriminating
variables, as well as the relative isolation of the lepton. Here, the
relative isolation is defined as the scalar \pt sum of all photons,
neutral and charged hadrons within an angular distance of $\Delta
R\equiv\sqrt{\smash[b]{(\Delta\phi)^2+(\Delta\eta)^2}}<0.3\,(0.4)$ for
electrons (muons), and divided by the \pt of the lepton candidate,
where $\phi$ is the azimuthal angle in radians. Loose (tight)
leptons are required to have $\pt > 10\,(20)\GeV$. For genuine electrons
(muons) originating from the primary vertex, the loose and tight
working points have efficiencies of approximately 90\% (98\%) and 70\%
(92\%), respectively. Corrections are applied to account for
contributions from charged and neutral pileup particles. Finally, events
containing tau leptons can contribute indirectly from leptonic decays
involving electrons or muons, which account for less than 3\% of the total 
signal cross section. 

Jets are clustered from PF candidates using the anti-\kt
algorithm~\cite{Cacciari:2008gp, Cacciari:2011ma} with a distance
parameter of 0.4. Two techniques are used, depending on the
data-taking period, to mitigate contributions from pileup. For Run 2
data, charged particles identified as having originated from pileup
vertices are discarded, and an offset correction is applied to correct
for remaining neutral
contributions~\cite{Sirunyan:2017ulk}. Additionally, for jets with
$\pt<50\GeV$, a multivariate pileup discriminant is applied,
using a tight working point~\cite{Sirunyan:2020foa}. For 2022 data,
the pileup-per-particle identification algorithm
(PUPPI)~\cite{Sirunyan:2020foa,Bertolini:2014bba} is used to mitigate
the effect of pileup at the reconstructed-particle level, making use
of local shape information, event pileup properties, and
tracking information.  Charged particles identified as having
originated from pileup vertices are discarded. The local information
is used to derive a weight encoding the probability that a neutral
candidate comes from a pileup vertex. Each candidate's four-momentum
is then scaled by this weight, creating PUPPI-weighted PF
candidates, before being clustered into jets, labelled as PUPPI
jets. For both methods, to adjust the average measured energy scale
of the jets to that of particle-level jets, corrections are derived
using simulated events, as a function of the reconstructed jet \pt and
$\eta$. In situ measurements of the momentum balance in dijet,
\phojets, \Zjets, and multijet events are used to determine any
residual differences between the jet energy scale in data and in
simulation, and appropriate corrections are
made~\cite{Khachatryan:2016kdb}. The energy resolution is also
corrected in simulation to improve the agreement with data. Jets are
selected in the range $\abs{\eta}< 2.4$ and with $\pt>20\GeV$. Jets
with an identified electron or muon within $\Delta R < 0.4$ are
rejected. Events with jets consistent with reconstruction
failures are vetoed.

The vector \ptvecmiss is computed as the negative
vector $\vec{p}_\mathrm{T}$ sum of all PF candidates in an event, and its
magnitude and azimuthal angle are denoted as \ptmiss and $\phi^{\text{miss}}$, respectively. For Run 2, \ptmiss is defined using PF candidates,
whereas, for Run 3, PUPPI-weighted PF candidates are utilised. Any correction
applied to individual objects is propagated correspondingly to the
\ptmiss~\cite{CMS:2019ctu}. Specific event filters have been designed to reduce
events with spurious \ptmiss arising from reconstruction failures or from genuine
showers not originating from the \pp collisions~\cite{CMS:2019ctu}.

The specific features of heavy-flavoured jets, in particular the
presence of displaced vertices, are used in a multivariate jet tagging
method. A working point of the \textsc{DeepJet}
algorithm~\cite{Bols:2020bkb} is used to tag \PQb quark jets with
${\pt > 20\GeV}$ and ${\abs{\eta} < 2.4}$, corresponding to a 90\%
efficiency for correctly identifying \PQb jets and a 40 (4)\%
probability of misidentifying a charm jet (light-flavour or gluon jet)
as a bottom jet.

\section{Event selection}
\label{sec:sel}

Events are selected in three steps. First, events are required to pass
criteria based on HLT algorithms that require the presence of at least
one electron (muon), satisfying isolation and identification
requirements, and trigger-level lepton \pt thresholds of 27--32
(24--27)\GeV. Exact threshold values depend on the data-taking
period. Second, a basic set of preselection criteria is applied that
remove reducible SM backgrounds, while retaining a high efficiency for
signal events. Third, a parameterised neural network
\cite{Baldi:2016fzo} is utilised to further separate signal from
backgrounds, and the output of the network is used to define the
signal region (SR).

\subsection{Preselection}

The final state contains a dilepton pair and two DM particles.  Based
on this, the preselection requires all events to contain at least two
tight leptons, with the leading lepton passing year-specific \pt
requirements, and a subleading lepton with $\pt > 20\GeV$. The leading
reconstructed lepton \pt thresholds are chosen to be $2\GeV$ above
the thresholds implemented in the lepton triggers, with values
outlined in Table~\ref{table:preselection}.

Each event is required to contain a dilepton pair---with no
constraints yet on the charges or flavours of the pair---that pass the
following requirements. As the signal leptons recoil against the DM
particles, the dilepton \pt is expected to be antiparallel with
$\ptvecmiss$, leading to the requirement $\abs{\Delta
\phi(\vec{p}_{\mathrm{T}}^{\Pell\Pell}, \ptvecmiss)} > 1\unit{rad}$. The recoil also
boosts the dilepton pair, motivating the requirement
$p_{\mathrm{T}}^{\,\Pell\Pell} > 15\GeV$, and the dilepton pair is more
collimated, hence the requirement $\cos(\Delta \phi_{\Pell\Pell}) >
-0.75$. The dilepton pair is required to have an invariant
mass of $12 < \mll < 80\GeV$. Finally, with little hadronic activity
expected in the signal, any event with two or more jets with
$\pt>30\GeV$ and $\abs{\eta}<2.4$ is vetoed.

\begin{table}[ht!]
  \centering
  \topcaption{Kinematic preselection of dilepton
      pairs. All events require at least one dilepton pair passing the
      "Dilepton" quantities, but with no constraints on the charges or
      flavours. The \pt of the leading lepton is dictated by the
      year-specific thresholds of the single-lepton triggers.
  }
    \renewcommand{\arraystretch}{1.3}
    \begin{tabular}{ll}
        Quantity                                                  & Requirement                                          \\
        \hline
        $N_{\Pell, {\text{tight}}}$                                                  & $\geq$2                                             \\
        Leading lepton \pt                                      & 2016: $>$26\GeV (muon), $>$29\GeV (electron) \\
                                                                  & 2017: $>$29\GeV (muon), $>$34\GeV (electron) \\
                                                                  & 2018: $>$26\GeV (muon), $>$32\GeV (electron) \\
                                                                  & 2022: $>$29\GeV (muon), $>$32\GeV (electron) \\
        Subleading lepton \pt                                   & $>$20\GeV                                        \\
        Dilepton invariant mass                                   & $12 < \mll < 80\GeV$                           \\
        Dilepton $\pt^{\Pell\Pell}$                               & $>$15\GeV                                        \\
        Dilepton $\abs{\Delta \phi(\vec{p}_{\mathrm{T}}^{\,\Pell\Pell}, \ptvecmiss)}$ & $> 1\unit{rad}$                                         \\
        Dilepton $\cos(\Delta \phi_{\Pell\Pell})$                 & $>$$-0.75$                                         \\
        Number of jets with $\pt>30\GeV$, $\abs{\eta}<2.4$            & $\leq$1                                             \\
    \end{tabular}
    \label{table:preselection}
\end{table}

\subsection{Variable definitions}

The following kinematic variables are particularly powerful for separating the signal from SM background processes with two leptons and \ptmiss in the final state. The transverse mass is defined as follows:
\begin{equation}
    \label{eq:mt}
    \mT(\vec{p}_{\mathrm{T}}^{\,\mathrm{a}}, \ptvecmiss) = \sqrt{2  \pt^{\,\mathrm{a}}  \ptmiss(1 - \cos[\Delta \phi(\vec{p}_{\mathrm{T}}^{\,\mathrm{a}}, \ptvecmiss)])},
\end{equation}
where $\vec{p}_{\mathrm{T}}^{\,\mathrm{a}}$ refers to the \pt of
the individual leptons. In this paper, we also consider a different
definition where $\vec{p}_{\mathrm{T}}^{\,\mathrm{a}}$ refers to the \pt of the dilepton system. In this case, the invariant mass of the
dilepton system can be included to obtain the transverse mass with
energy, $\mT^E$, defined as:
\begin{equation}
  \label{eq:mte}
    \mT^E(\vec{p}_{\mathrm{T}}^{\,\Pell\Pell}, \mll, \ptvecmiss) = \sqrt{2  \sqrt{(\pt^{\,\Pell\Pell})^2 + \mll^2}  \ptmiss (1 - \cos[\Delta \phi(\vec{p}_{\mathrm{T}}^{\,\Pell\Pell}, \ptvecmiss)])}.
\end{equation}
The stransverse mass (\mTii)~\cite{MT2,MT2_ref2,MT2_ref3} is the generalisation of the \mT for the case where two visible and two non-interacting
particles are produced. The \pt of the two visible particles is given by
$\pt^{\,\mathrm{a}}$ and $\pt^{\,\mathrm{b}}$. In this case, the \pt of the two
invisible particles, $\vec{q}_{\mathrm{T}}^{\,\mathrm{1}}$ and $\vec{q}_{\mathrm{T}}^{\,\mathrm{2}}$, are
unknown, but their sum is constrained by $\vec{q}_{\mathrm{T}}^{\,\mathrm{1}} +
\vec{q}_{\mathrm{T}}^{\,\mathrm{2}} = \ptvecmiss$. Furthermore, to include the case where the
non-interacting particles have mass $\chi$, \mTii uses a
generalised version of the transverse mass, $\mT'$:
\begin{equation}
    \label{eq:mt'}
    \mT'(\vec{p}_{\mathrm{T}}^{\,\mathrm{a}}, \ptvecmiss, \chi) = \sqrt{2  \pt^{\,\mathrm{a}}  \sqrt{(\ptmiss)^2 + \chi^2 } \left(1 - \cos[\Delta \phi(\vec{p}_{\mathrm{T}}^{\,\mathrm{a}}, \ptvecmiss)]\right)},
\end{equation}
The definition of \mTii is then:
\begin{equation}
  \label{eq:mtii}
    \mTii (\vec{p}_{\mathrm{T}}^{\,\mathrm{a}},\vec{p}_{\mathrm{T}}^{\,\mathrm{b}},\ptvecmiss,\chi) = \min_{\vec{q}_{\mathrm{T}}^{\,\mathrm{1}} + \vec{q}_{\mathrm{T}}^{\,\mathrm{2}} =  \ptvecmiss} \left[ \max\left\{ \mT'(\vec{p}_{\mathrm{T}}^{\,\mathrm{a}},  \vec{q}_{\mathrm{T}}^{\,\mathrm{1}}, \chi), \mT'(\vec{p}_{\mathrm{T}}^{\,\mathrm{b}},  \vec{q}_{\mathrm{T}}^{\,\mathrm{2}}, \chi)  \right\} \right],
\end{equation}
where "$\min$" is a minimisation with the constraint
$\vec{q}_{\mathrm{T}}^{\,\mathrm{1}} + \vec{q}_{\mathrm{T}}^{\,\mathrm{2}} =
\ptvecmiss$, and $\max \{ \mT'(\dots), \mT'(\dots) \}$ returns
the maximum of the two $\mT'$ calculations. To use $\mTii$ for its
discriminatory power, the variable is calculated under two mass
hypotheses. The first is with $\chi = 0\GeV$, which corresponds to the
SM-like case where the invisible particles are massless neutrinos. The
second is $\chi = 80\GeV$, which is the IDM-like case where the
invisible particles, $\PH$, are massive. Different choices of values
for $\chi$, \eg matching the generated signal \mh, do not impact the
signal vs. background discrimination power.

Finally, the jet-Z balance (JZB)~\cite{CMS:2012iqa} is defined as:
\begin{equation}
  \label{eq:jzb}
    \mathrm{JZB} = \abs{\ptvecmiss + \vec{p}_{\mathrm{T}}^{\,\Pell\Pell}} - \abs{\vec{p}_{\mathrm{T}}^{\,\Pell\Pell}}.
\end{equation}

For SM \DYjets events, where the \ptmiss stems
from detector effects rather than the underlying physical process,
this quantity is expected to be symmetric around 0. Whereas, for SM
events with neutrinos, or signal events, the dilepton \pt and \ptmiss are correlated, leading to a
JZB distribution skewed towards more negative values, allowing
discrimination against \DYjets events.

\subsection{Signal region selection using a parameterised neural network}
\label{sec:pnn}

With the preselection defined in Table~\ref{table:preselection}, an
additional selection is applied specifically targeting the signal
topology. Every event is required to have an OCSF dilepton pair, with
both leptons passing the tight working point ($N_{\Pell,
  {\text{tight}}}=$\;2). A veto is applied on any event with an
additional loose lepton. Finally, to reduce the background from
$\ttbar$ and $\PQt\PW$ events, a veto on any event containing a
\PQb-tagged jet is applied.

\begin{figure}[bh!]
    \centering
    \includegraphics[width=0.49\textwidth]{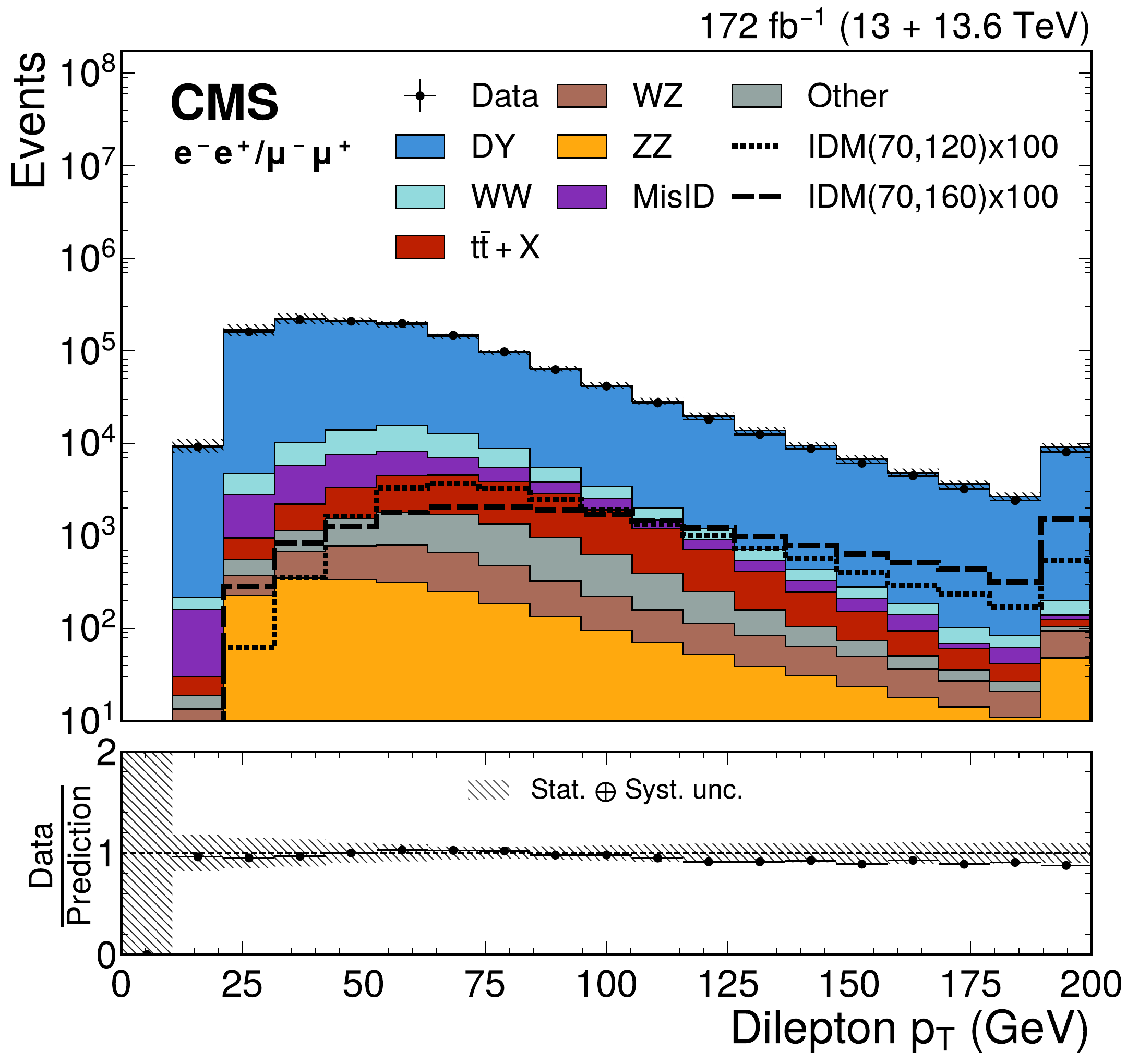}\hfill
    \includegraphics[width=0.49\textwidth]{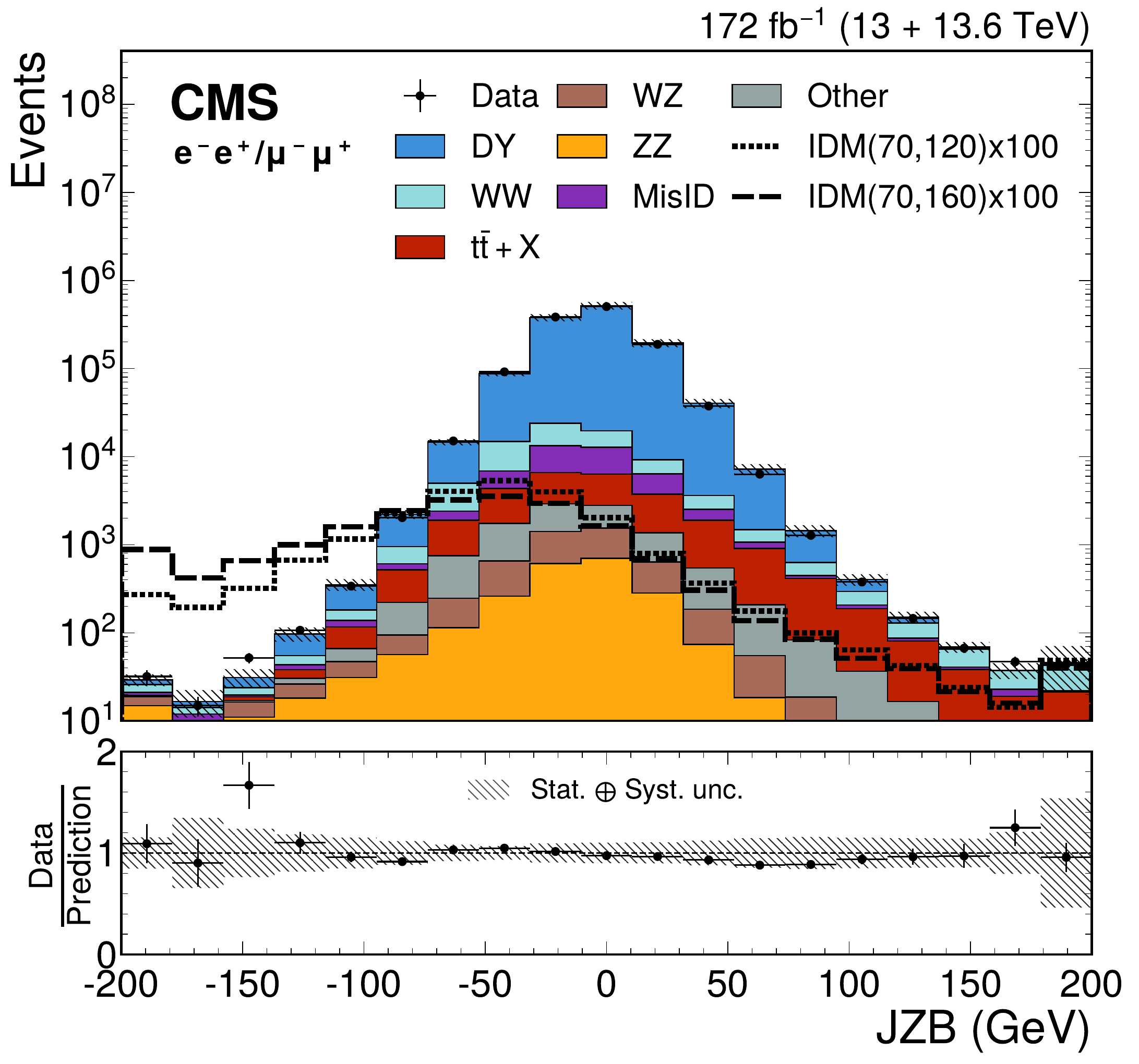} \\
    \includegraphics[width=0.49\textwidth]{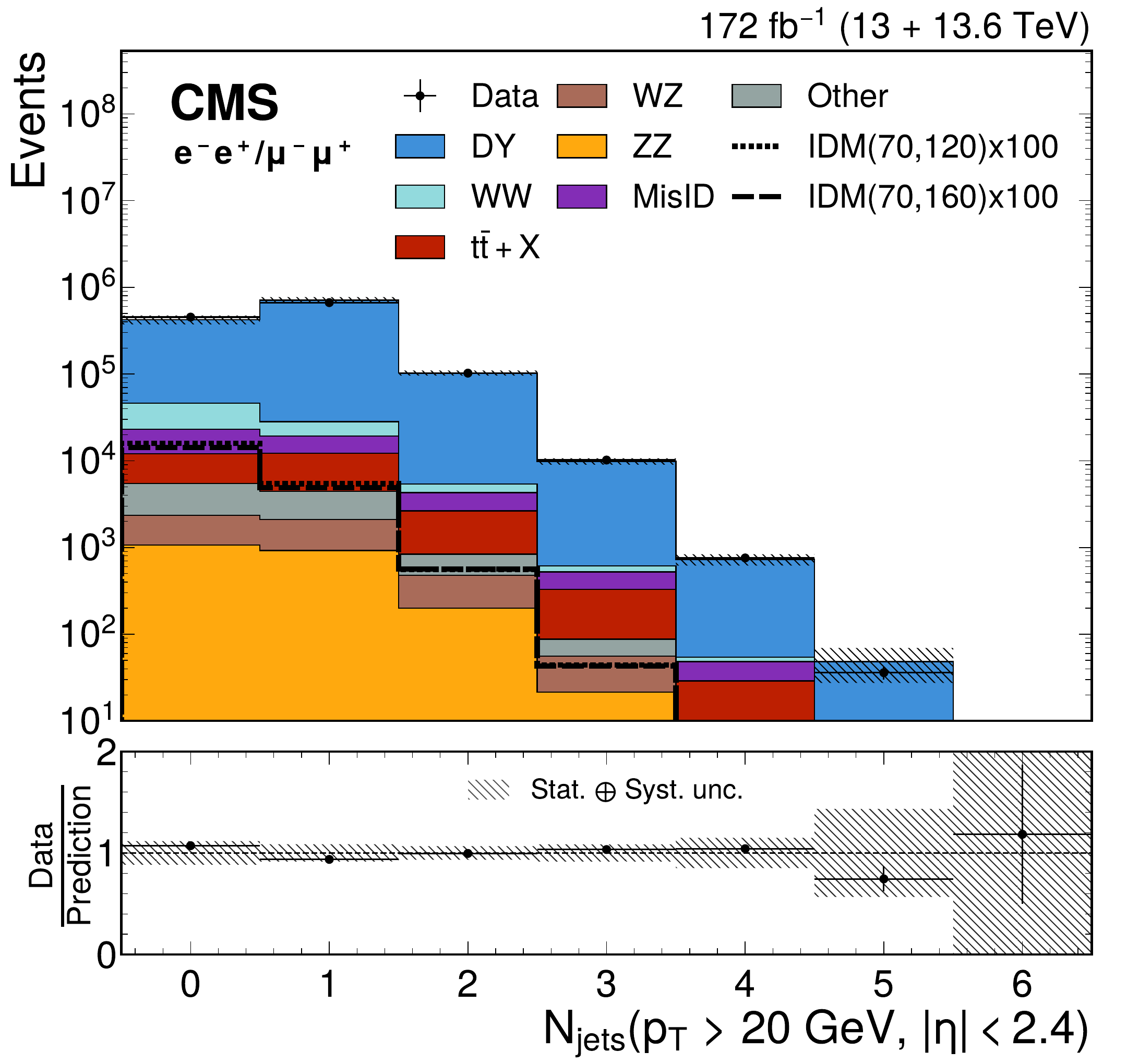}\hfill
    \includegraphics[width=0.49\textwidth]{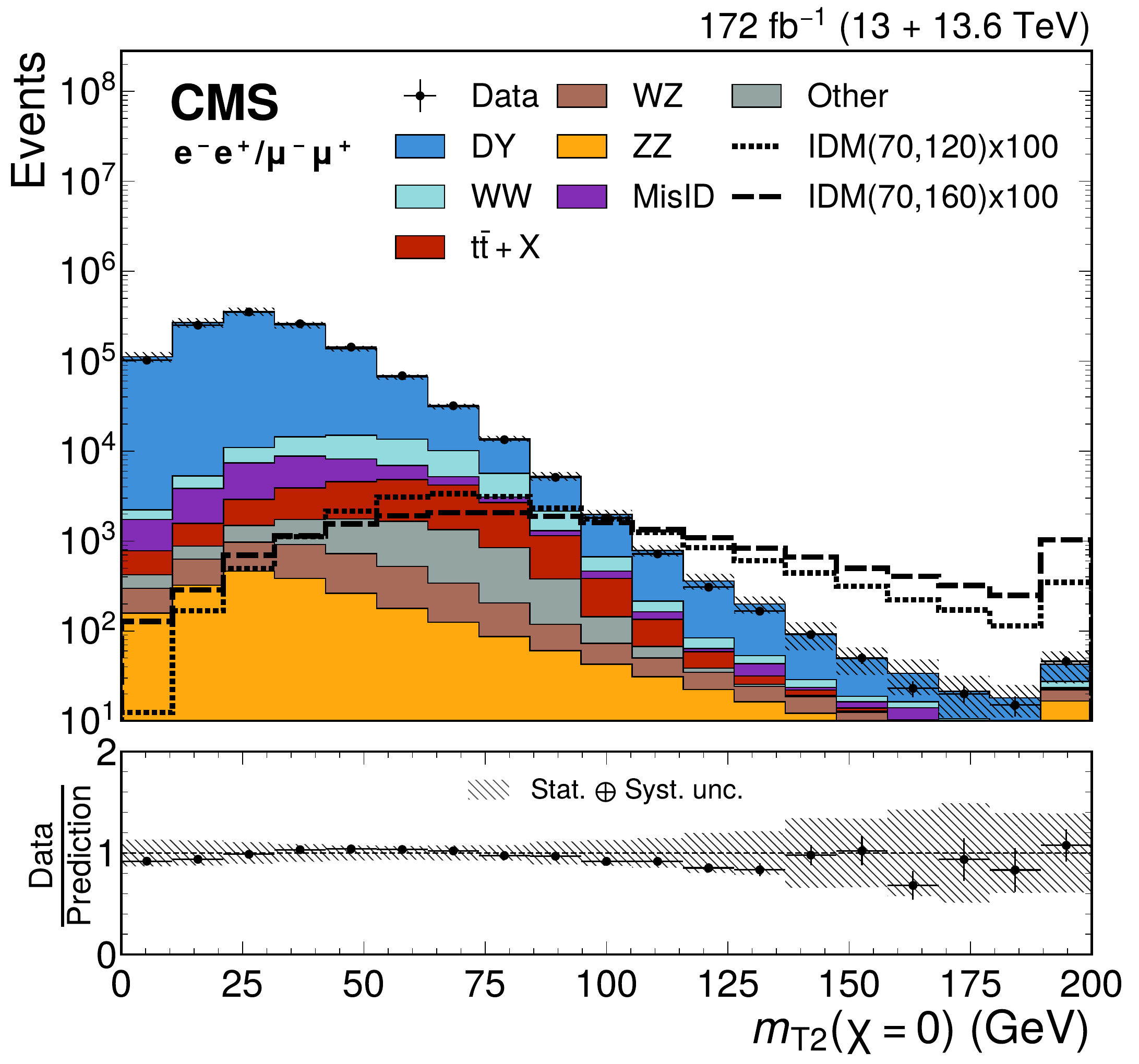}
    \caption{Distributions of key variables after the preselection
        and OCSF pair selection for 2016--2018 and 2022 combined. All SM
        processes are modelled with simulation, except the contribution
        from events with jets misidentified as leptons (MisID), which are
        calculated using control samples in data, as described in Section
        \ref{sec:misid}. Two representative IDM signal samples are also shown,
        indicated by IDM(\mh,\ma) with the masses given in \GeV, and
        with their normalisation scaled by a factor of 100 for
        clarity. The first and last bins contain the underflow and overflow
        events, respectively. The lower panels show the ratio of the
        data over the SM prediction. The error bars show the
        data statistical uncertainties, while the hatched bands include both
        MC statistical and systematic components.}
    \label{fig:preselections}
\end{figure}

Example distributions after this selection
are shown in Fig.~\ref{fig:preselections} for Run~2 and 2022
combined. The data distributions are compared to the distributions of
all SM processes, which are modelled with simulation, except for
events with a misidentified electron or muon, denoted MisID, which are
calculated using the method described in Section \ref{sec:misid}. The
contribution labelled "Other" contains events from single top and
triboson processes. In addition, two representative IDM signal points
are overlaid, with their normalisations multiplied by a factor of 100
for clarity.  Generally, the data agree with the SM background
prediction within the quoted uncertainties, which contain both
statistical and systematic contributions detailed in
Section~\ref{sec:systs}.

Signal events have a broad range of possible topologies depending on
the masses $(\mh, \ma)$. To ensure a good signal-background separation
across all points, a parameterised neural network (pNN) is
used~\cite{Baldi:2016fzo}. Parameterised neural networks extend normal
neural networks by also including the signal parameters $(\mh, \ma)$
as input, resulting in the network learning how to discriminate signal
versus background depending on the values of $(\mh,\ma)$. As a
consequence, only a single network, which can achieve optimal
performance for each point, is required for the entire parameter
space. Furthermore, given that pNN outputs vary smoothly as functions
of $(\mh,\ma)$, it is possible to interpolate between the simulated
signal points, thus ensuring strong sensitivity over the entire
parameter space.

To reduce the complexity of the pNN model, as well as reduce training
times, the choice of input features is optimised using boosted
decision trees (BDT) in an iterative process. The process starts with
a large set of inputs involving both basic kinematic variables and
derived variables. Subsequently, for each individual signal point, an
\textsc{XGBoost} BDT is trained using all inputs~\cite{xgboost}. To quantify
the importance of each feature, the \textsc{XGBoost} "Gain" measure is
utilised, which corresponds to the average improvement to the loss
score each feature provides over the whole tree. The least important
feature is removed, the BDT is trained again, and the process is
repeated until the signal efficiency, at a background efficiency of
0.1\% found by cutting on the BDT output score, reduces by more than
1\%. This background efficiency is chosen as it approximately
corresponds to the background efficiency in the most sensitive regions
of the analysis. As the kinematic distributions vary with the signal
points, leading to different variables being selected, the
optimisation procedure is performed for all points, and the final set
of features is the union of the features selected for each
point. Furthermore, to take into account the fact that a NN can
"derive" variables, all the basic kinematic variables are kept
independently of their impact. After the procedure, a total of 26
features, 14 basic variables and 12 derived variables, are selected:

\begin{itemize}
  \item leading/subleading lepton \pt, $\eta$, $\phi$;
  \item leading/subleading jet \pt, $\eta$, $\phi$;
  \item $\ptmiss$, $\phi^{\text{miss}}$;
  \item dilepton $\pt^{\,\Pell\Pell}$, $\eta^{\,\Pell\Pell}$,
        $\phi^{\,\Pell\Pell}$, $\mll, \Delta R_{\Pell\Pell}$;
  \item JZB (Eq.~(\ref{eq:jzb}));
  \item transverse mass, $\mT(\vec{p}_{\mathrm{T}}^{\,\Pell\Pell}, \ptvecmiss)$ (Eq.~(\ref{eq:mt}));
  \item transverse mass with energy, $\mT^E(\vec{p}_{\mathrm{T}}^{\,\Pell\Pell}, \ptvecmiss)$ (Eq.~(\ref{eq:mte}));
  \item stransverse masses, $\mTii
          (\vec{p}_{\mathrm{T}}^{\,\Pell_1},\vec{p}_{\mathrm{T}}^{\,\Pell_2},\ptvecmiss,\chi=0,80\GeV)$ (Eq.~(\ref{eq:mtii}));
  \item $\Delta \phi (j_1, \ptvecmiss)$;
  \item $\Delta \phi (j_1, \vec{p}_{\mathrm{T}}^{\,\Pell\Pell})$.
\end{itemize}

Each of the simulated samples are split into three sets: training, validation, and
test with fractions of the total being 50\%, 20\%, and 30\%, respectively.
The training data set is used to optimise the pNN parameters, and the validation
set is used to measure the performance of the model during training. Finally,
the test set is used to evaluate the performance of the model, to 
ensure that there is no overtraining, and to obtain the results of the analysis. 
Events are initially weighted according
to their relative expected yields. Before training, the background samples are
reweighted such that the total sum of background weights equals the total sum
of all signal weights. The signal samples are also reweighted such that the sum
of weights for each sample is equal. During training, the signal samples are
passed through the pNN with the IDM parameters $(\mh, \ma)$ at which they were
simulated. The background samples are randomly assigned $(\mh, \ma)$ values
from the set of signal $(\mh, \ma)$ values used in the training.

The pNN is implemented using \textsc{PyTorch}~\cite{pytorch}, and consists of a
feedforward network with six layers, each layer containing 60 hidden units and
an ELU activation function~\cite{Clevert2015FastAA}, where the architecture was
determined over a grid search of different hyperparameters. The model is
trained using the Adam optimiser~\cite{adamOptimiser} with a learning rate of
$10^{-4}$ to minimise the weighted binary cross-entropy loss
function~\cite{Terven2025LossFunctions}. The pNN performance is observed to be
consistent with individual \textsc{XGBoost} BDTs trained separately on each signal
point.

The SR is defined using the pNN output, leveraging the fact that the
signal peaks towards an output of 1. Ideally, separate binning could be used
for each signal, optimising the sensitivity for every point, however this leads
to ambiguity when interpolating between points. Instead, a single set of bins
is used for the entire parameter space, but with separate binning for each
channel, \ee and \mumu. Despite using the same binning, the loss in sensitivity
was observed to be less than 1\%. To ensure that the distributions remain
smooth, each bin is required to have a maximum of 10\% statistical uncertainty
in the MC predictions. As the signal distributions peak towards a pNN value of
1, the optimisation procedure starts at 1 and then uses a sliding window towards
lower pNN values. If a bin has a background statistical uncertainty of less
than 10\% for all signal hypotheses then it is accepted, and the sliding window
starts again from the previous bin. This is repeated down to a pNN value of
0.9, which approximately corresponds to a minimum signal efficiency of 50\%.
The SR selection is summarised in Table~\ref{table:signal_selections}.
For all signal mass points, the dominant signal contribution ($>$99\%) in the SR is from the
$\PH\PH\Pell\Pell$ final state, where $\Pell=\Pe,\PGm,$ or $\PGt$.

\begin{table}[ht!]
    \centering
    \topcaption{Signal region selection. Events must also pass the preselection outlined in Table~\ref{table:preselection}.
        A veto is applied on any event with additional loose leptons.}
    \renewcommand{\arraystretch}{1.2}
    \begin{tabular}{ll}
        Quantity                      & Requirement               \\
        \hline
        $N_{\Pell, {\text{tight}}}$    & $=$2                     \\
        Dilepton pair  flavour/charge & $\Pep\Pem/\PGmp\PGmm$ \\
        Number of \PQb jets           & $=$0                     \\
        pNN output           & $>$0.9                     \\
    \end{tabular}
    \label{table:signal_selections}
\end{table}

To interpolate between different mass points for the simulated
background processes and data, the distributions are calculated by
evaluating the pNN on these samples with the desired mass hypothesis,
$(\mh,\ma)$. The signal samples, however, are simulated at specific
mass points, and as such can only give the signal prediction for that
point. Therefore, to find the expected signal distributions for the
intermediate mass points, cubic spline interpolation is used for each
bin separately, with the shape parameter, $\epsilon$, determined by
minimising the leave-one-out cross-validation
error~\cite{shapeParameterAlgorithm} across all mass points. To
quantify the performance of the interpolation, the validation points
in Fig.~\ref{fig:parameter_scan_mH_mA-mH} are utilised. The
interpolated distributions are compared to the true distributions
calculated using samples simulated at the validation points, and are
observed to be consistent within the statistical uncertainties.

\section{Background estimation}
\label{sec:bkg}

Backgrounds are estimated using a combination of simulated samples and
techniques based on control samples in data. Utilising the latter
leads to improved sensitivity compared to a simulation-only approach
by providing tighter constraints on the dominant background
normalisations and reducing the associated systematic
uncertainties. After the preselection and signal selection outlined in
Section~\ref{sec:sel}, the largest background is $\DYjets$. After the
requirement of $\pnn > 0.9$, however, this background source is
significantly reduced, with the largest remaining background sources
in the most significant bin (close to 1) being $\PW\PW$, $\ttbar$,
$\PZ\PZ$, and $\PW\PZ$. For the estimation of the backgrounds from
data, two variables are used, namely the pNN output for the
discrimination of the signal against each background source, and the
dilepton \pt, which allows to better discriminate between the
$\ttbar$ and $\PW\PW$ components (see Section~\ref{sec:wwtt}). For
each of the main background processes, their shape is determined with
simulation, while their normalisation is constrained with data using
CRs enriched in each process. An additional source of the background
in the SR and CRs is {\PW}+jets and QCD multijet events where one or
more jets are misidentified as leptons.  These are estimated using a
misidentification background determination method based on control
samples in data. An overview of the CR selections is shown in
Table~\ref{table:cr_selections}, and a description of each CR is given
in this section. Other minor backgrounds (from \DYjets, single top quark
and $\PV\PV\PV$ processes) are estimated directly using the predictions from
simulations.

\begin{table}[ht!]
    \centering \topcaption{Selections for the control regions.
        For all events, we require at least one dilepton pair passing the
        preselection outlined in Table~\ref{table:preselection}, as well
        as any additional requirements given in this table. The second
        dilepton in the ZZ CR does not have to pass the selection in
        Table~\ref{table:preselection}. The $N_\Pell$ ($N_{\Pell, {\text{tight}}}$) 
        selection indicates the total number of loose leptons selected and the 
        number of those that must also pass the tight selection criteria. 
        All regions have a veto on any
        additional loose leptons. The fit variable refers to the fit
        procedure described in Section~\ref{sec:res}.  }
    \renewcommand{\arraystretch}{1.2}
    \begin{tabular}{l l l}
        Control region                                              & Quantity                              & Requirement                       \\
        \hline
                    \multicolumn{3}{l}{Fit variable: Dilepton \pt}                    \\
\hline
        \rule{0pt}{2.7ex}    \multirow{5}{*}{$\PWm\PWp/\ttbar$ }    & $N_\Pell$ ($N_{\Pell, {\text{tight}}}$) & $=$2 (2)                         \\
                                                                    & Dilepton   flavour/charge             & $\Pepm\PGmmp$                 \\
                                                                    & Number of \PQb jets                   & $=$0                             \\
                                                                    & $\pnn$                                & $>$0.9                           \\
        \hline
        \rule{0pt}{2.7ex} \multirow{5}{*}{$\ttbar$ }                & $N_\Pell$ ($N_{\Pell, {\text{tight}}}$) & $=$2 (2)                         \\
                                                                    & Dilepton   flavour/charge             & $\Pepm\PGmmp$                 \\
                                                                    & Number of \PQb jets                   & $\geq$1                          \\
                                                                    & $\pnn$                                & $>$0.9                           \\
        \hline
        \rule{0pt}{2.7ex} \multirow{5}{*}{MisID different-flavour}   & $N_\Pell$ ($N_{\Pell, {\text{tight}}}$) & $=$2 (2)                         \\
                                                                    & Dilepton  flavour/charge              & $\Pepm\PGmpm$                 \\
                                                                    & Number of \PQb jets                   & $=$0                             \\
                                                                    & $\pnn$                                & $>$0.9                           \\
        \hline
                   \multicolumn{3}{l}{Fit variable: $\pnn$}                    \\
        \hline
        \rule{0pt}{2.7ex} \multirow{5}{*}{$\PZ\PZ$ }                & $N_\Pell$ ($N_{\Pell, {\text{tight}}}$) & $=$4 ($\geq$2)                  \\
                                                                    & Both dilepton flavour/charge          & $\Pep\Pem/\PGmp\PGmm$         \\
                                                                    & Second dilepton  mass                 & $\abs{\mll-m_{\PZ}}<10\GeV$           \\
                                                                    & $\pnn$                                & $>$0.8                           \\
        \hline
        \rule{0pt}{2.7ex} \multirow{4}{*}{$\PW\PZ$ opposite-charge} & $N_\Pell$ ($N_{\Pell, {\text{tight}}}$) & $=$3 (3)                         \\
                                                                    & Dilepton  flavour/charge              & $\Pepm\Pemp/\PGmpm\PGmmp$ \\
                                                                    & $\pnn$                                & $>$0.9                           \\
        \hline
        \rule{0pt}{2.7ex} \multirow{4}{*}{$\PW\PZ$ same-charge}     & $N_\Pell$ ($N_{\Pell, {\text{tight}}}$) & $=$3 (3)                         \\
                                                                    & Dilepton  flavour/charge              & $\Pepm\Pepm/\PGmpm\PGmpm$ \\
                                                                    & $\pnn$                                & $>$0.9                           \\
        \hline
        \rule{0pt}{2.7ex} \multirow{5}{*}{MisID same-flavour}       & $N_\Pell$ ($N_{\Pell, {\text{tight}}}$) & $=$2 (2)                         \\
                                                                    & Dilepton  flavour/charge              & $\Pepm\Pepm/\PGmpm\PGmpm$ \\
                                                                    & Number of \PQb jets                   & $=$0                             \\
                                                                    & $\pnn$                                & $>$0.9                           \\
    \end{tabular}
    \label{table:cr_selections}
\end{table}

\subsection[ZZ CR]{The \texorpdfstring{$\PZ\PZ$}{ZZ} control region}

The $\PZ\PZ$ process constitutes the largest background, stemming from
one of the Z bosons decaying to an OCSF lepton pair, and the other Z
boson decaying to neutrinos. This can be emulated with a four-lepton
selection containing two OCSF dilepton pairs~\cite{CMS:2020ulv}. One
of the OCSF dilepton pairs is required to pass the preselection shown
in Table~\ref{table:preselection}. The other dilepton pair is removed
from the event and its \pt is added vectorially to the \ptmiss,
emulating the neutrino contribution. Furthermore, the dilepton pair
that acts as the neutrino pair is required to only pass the loose
lepton selection, and have $\abs{\mll-m_{\PZ}}<10\GeV$ to remove
$\gamma^* \to \Pell \Pell$ contributions. For events with four leptons
of the same flavour ($\ee\ee/\mumu\mumu$) there is an ambiguity in
pairing the leptons. In this case, the pair with an invariant mass
closest to the \PZ boson is chosen as the candidate neutrino
pair. Each event requires four loose leptons, at least two of which
are tight. Events with more than four loose leptons are vetoed.

After four-lepton events are selected, the $\PZ\PZ$ CR
requires a looser selection on the pNN output than in the SR,
$\pnn > 0.8$, to increase the number of events. The $\PZ\PZ$
normalisation was found to not depend strongly on the pNN threshold,
given the large statistical uncertainty of the data sample.  Finally,
the pNN output score is used as the observable in the fit when
extracting a potential signal.

\subsection[WWTT CR]{The \texorpdfstring{$\PW\PW$/$\ttbar$}{WW/ttbar} control region}
\label{sec:wwtt}

For $\PW\PW$ and $\ttbar$ production, events with different-flavour leptons are
produced in equal amount to events with same-flavour leptons, excluding
reconstruction-specific effects. This symmetry is utilised to define two CRs to
constrain the $\PW\PW$ and $\ttbar$ background normalisations.

Each event is required to have exactly two tight opposite-charge
different-flavour leptons, $\Pepm\PGmmp$, and must pass the
preselection shown in Table~\ref{table:preselection}. Following this,
each event is required to have $\pnn > 0.9$ to match the SR phase
space. The events are further split into those with no \PQb jets,
$N_{\text{\PQb jets}} = 0$, resulting in a $\PW\PW$/$\ttbar$ CR, and
those with $N_{\text{\PQb jets}} \geq 1$, giving a $\ttbar$-dominated
CR. For both CRs, when extracting a potential signal, instead of
fitting to the pNN output like in the $\PZ\PZ$ CR, the dilepton \pt
distribution is used as this provides better discrimination between
the $\PWm\PWp$ and $\ttbar$ processes. 
The selected $\ttbar$ events typically have a greater dilepton \pt than those from the $\PWm\PWp$ process. 
To calculate the optimal
binning, a sliding window method is applied---starting from low \pt
and scanning towards higher \pt---which accepts a bin if the
statistical uncertainties of the $\PWm\PWp$ and $\ttbar$ processes are
both below 10\%. Signal events can enter the CR through the
$\PH\PH\PGt\PGt$ or $\PH\PH\Pell\Pell\PGn\PAGn$ final states, however, after
the requirement $\pnn > 0.9$ this contribution is negligible.

\subsection[WZ CR]{The \texorpdfstring{$\PW\PZ$}{WZ} control region}
\label{sect:WZEstimation}

The $\PW\PZ$ contribution entering the SR comes primarily from the
fully leptonic final state, with negligible contributions from lepton+jets
and fully-hadronic decays. These can contribute in two ways. First, when the
leptons from the decay of the \PZ boson form the signal dilepton pair, and the
lepton from the \PW boson decay falls out of acceptance. Second, when the
dilepton pair is formed from one lepton from the \PZ boson decay and with the
lepton from the \PW boson decay, and the remaining lepton from the \PZ boson
falls out of acceptance. These are both estimated using a three-lepton
selection.

The first $\PW\PZ$ CR requires exactly three tight leptons, containing an OCSF
dilepton pair, which passes the preselection in Table~\ref{table:preselection}.
Events with an additional loose lepton are vetoed. For events with three
leptons of the same flavour ($\Pe\Pe\Pe, \PGm\PGm\PGm$) the ambiguity in
identifying the pair coming from the \PZ boson is resolved by choosing the pair
that produces an invariant mass closest to the $\PZ$ boson mass. Subsequently,
events are required to pass $\pnn > 0.9$ to match the SR phase
space. The \ee and \mumu channels are combined into one CR, keeping the
data-taking years separated, and the pNN output is used as the observable in
the fit when extracting a potential signal. The second $\PW\PZ$ CR, targeting
the case where one lepton is from the \PZ boson decay and the other from the
\PW boson decay, has identical requirements to the first, but requires a same-charge
 same-flavour dilepton pair.

\subsection{Estimates of backgrounds with misidentified leptons}
\label{sec:misid}

Background processes that contain jets can enter the different selections when
one or more of the jets are incorrectly reconstructed as a lepton. This
contribution is referred to as MisID events, and is estimated using the
"misidentification method" from Ref.~\cite{MissID1}.

This method uses the misidentification rate, which is defined as the
probability that a jet, that has already been reconstructed as a loose lepton,
also passes the tight lepton requirements. The misidentification rate is
measured using a sample enriched in {\PZ}+jets events. Each event
requires two tight leptons with $\pt > 20\GeV$ that form an OCSF pair which
satisfy $\abs{\mll-m_Z}< 15\GeV$. A third loose lepton (electron or muon)
is also required, with $\pt > 20\GeV$, which acts as the probe lepton for
measuring the misidentification rates. To suppress the contributions from
diboson events with neutrinos in the final state, all events must pass
$\ptmiss<25\GeV$. Events with a fourth loose lepton are vetoed to reduce the
contribution from the $\PZ\PZ \to 4\Pell$ process. Contributions from
events with prompt third leptons are subtracted using estimates from
simulation. The misidentification rate is estimated in bins of the loose lepton
\pt and $\eta$, separately for electrons and muons.

The MisID estimate is then calculated using data events with the same selection
as the SR or CRs, but where one or more lepton fails
the tight lepton requirements. In the case of the SR and
the $\PW\PW/\ttbar$ CR, events where one lepton fails are predominantly
{\PW}+jets events, and those where both fail are primarily QCD multijet
events. Using the misidentification rates, the data events are weighted to
calculate the expected number of MisID events with all leptons passing the
tight selection. Contamination from prompt events is subtracted using
simulation.

To constrain the normalisations of the MisID estimates, dedicated MisID CRs
are defined, exploiting the fact that jets can be misidentified as
leptons of either charge. The first MisID CR requires events with exactly two
tight leptons that form a same-charge same-flavour dilepton pair,
$\Pepm\Pepm/\PGmpm\PGmpm$, which passes the preselection in
Table~\ref{table:preselection}. Events with an additional loose lepton are vetoed, as
are those with a \PQb-tagged jet. Each event must then pass $\pnn > 0.9$ to match the
SR phase space, and then the pNN score is used as the variable in
the fit when extracting the signal. In addition, another MisID CR is defined,
which has the same requirements as the first, but instead demands the dilepton
pair to be same-charge and different-flavour, $\Pepm\PGmpm$. In this case,
the variable used in the fit is the dilepton \pt.

\section{Systematic uncertainties}
\label{sec:systs}

Several sources of experimental and theoretical uncertainties are considered in
this analysis, affecting either the total yield of a process, or the yield and
shape of the distributions. An overview of the systematic uncertainties and
their correlations is given in the following.

\begin{description}
    \item[Lepton efficiencies] The trigger, reconstruction, and identification
        efficiencies of electrons and muons have been measured in simulation, and
        corrected to match the observed efficiencies in data from \DYjets events,
        parameterised as a function of lepton \pt and
        $\eta$~\cite{CMS:2020uim,CMS:2018rym,Sirunyan:2017ezt}. The corrections are
        derived independently for the selection criteria of interest. The uncertainties
        associated with the corrections are propagated to the likelihood model. Except
        for the lepton trigger efficiencies, the lepton uncertainties are assumed
        correlated across years. However, electron and muon corrections are
        uncorrelated with each other, and when different working points are chosen, the
        corresponding uncertainties are kept uncorrelated.

    \item[Jet calibration] The effects of the jet energy scale (JES) and jet energy
        resolution (JER) uncertainties are studied explicitly by varying them by one
        standard deviation according to the source of uncertainty, and propagating the
        changes to $\ptmiss$~\cite{Khachatryan:2016kdb}. For the signal and the minor
        backgrounds, the impact on the simulated yields is studied in the SR as
        functions of the pNN output, with a fit procedure to remove the statistical
        contributions in the less-populated bins. In total, 27 JES uncertainty sources
        are considered for each data-taking year, with partial correlations between the
        different data sets for Run 2. The JER uncertainties are uncorrelated between
        all data-taking years. For Run 2, there is an additional uncertainty in the jet
        pileup identification efficiency.

    \item[Jet flavour tagging] The efficiency of the \PQb-tagging
      algorithm is measured in simulation, and corrected to match the
      efficiencies observed in data. These are parameterised as
      functions of the jet \pt, $\eta$ and generator-level
      flavour~\cite{Bols:2020bkb}. This uncertainty is partially
      correlated across data-taking years.

    \item[Pileup reweighting] Simulated events are weighted to match the distribution of
        the number of interactions to the distribution measured in the data. To model the
        uncertainty associated with this procedure, the total inelastic \pp cross
        section is varied by $\pm$5\%~\cite{CMS:2018mlc}. This uncertainty is
        correlated across data sets.

    \item[Level-1 trigger inefficiency] During the 2016 and 2017 data-taking
      periods, partial mistiming of signals in the forward region of
      the ECAL endcaps (${2.5<\abs{\eta}<3.0}$) led to a 
      reduction in the L1 trigger
      efficiency~\cite{Sirunyan:2020zal}. A separate correction was
      determined using an unbiased data sample and applied to
      simulated events to reproduce the loss of efficiency. The
      uncertainty in the correction is propagated through the
      analysis, leading to a maximum 2\% effect on the pNN output
      shape. This problem was resolved before the 2018 data-taking
      period.

    \item[Integrated luminosity] The uncertainty in the integrated luminosity is
        separated into correlated and uncorrelated contributions across data-taking
        years~\cite{CMS:2021xjt,CMS:2018elu,CMS:2019jhq,CMS-PAS-LUM-22-001}. Eight
        independent sources are identified to take into account the correlations across
        Run 2 data sets: an uncorrelated uncertainty of 1.0\%, 2.0\% and 1.5\% for 2016, 2017,
        and 2018, respectively; an uncertainty correlated across all years of 0.6\%, 0.9\%
        and 2.0\% for 2016, 2017 and 2018, respectively; additional correlated
        uncertainties of 0.6\% and 0.2\% for 2017 and 2018, respectively. For Run 3, the
        total uncertainty is 1.4\% in 2022. These uncertainties do not affect the
        \ttbar, $\PWm\PWp$, $\PW\PZ$, and $\PZ\PZ$ backgrounds which are estimated from
        data.

    \item[Drell--Yan modelling] No dedicated CR is defined for \DYjets, which
        provides a subdominant contribution in the most sensitive bins of the signal
        region. At lower values of pNN output, however, the \DYjets process becomes
        more significant, particularly for lower mass splitting signal samples. The
        modelling of this background is checked using a sideband region, which follows
        identical selections to the SR, but instead requires $0.8<\pnn<0.9$.
        A simultaneous fit to this region and to all other CRs is performed
        to extract a normalisation factor for \DYjets events. This
        normalisation factor is found to be generally around 1.25, with an absolute
        uncertainty of approximately 0.25. Instead of applying this factor directly to
        the SR contribution from the \DYjets process, a conservative
        uncertainty of 75\% is assigned to the \DYjets yields predicted from
        simulation, and included as a nuisance parameter in the statistical model. In
        addition, to cover the uncertainty in the shape of the \DYjets background, a
        theory systematic uncertainty is defined by using no
        corrections, or twice the NLO/LO factor corrections. All Drell--Yan uncertainties are
        uncorrelated across the data-taking years.

    \item[Parton distribution functions] The uncertainty from the
      choice of the PDFs is estimated following the PDF4LHC
      recommendations for Run 2 \cite{PDFrecommendationsRun2} and Run
      3 \cite{PDFrecommendationsRun3}, and embedded into the produced
      samples. These uncertainties are propagated using either the
      Hessian method or the MC method, depending on the choice of generator.
      This uncertainty is correlated across data sets.

    \item[Renormalisation and factorisation scales] Theoretical
      uncertainties are included to account for the truncation of the
      perturbative ME calculations, with higher-order terms not
      included. Each scale is varied independently by factors of 2 and
      0.5, and the resulting uncertainties are split into uncorrelated
      systematic uncertainties for $\PV,\PV\PV,\PV\PV\PV$, and
      top-quark processes. This uncertainty is correlated across data sets.

    \item[MC statistical uncertainties] The limited size of the simulated samples results
        in statistical fluctuations around the nominal prediction. The bin-by-bin
        uncertainty is accounted for using the Barlow--Beeston method
        \cite{MCstatUncert}. As described in Section~\ref{sec:pnn}, the uncertainty in
        the signal interpolation is covered by the statistical uncertainty of the
        samples. For any interpolated signal distribution, the bin-by-bin statistical
        uncertainty is given by the average uncertainty of the neighbouring simulated
        mass points.

\end{description}

\section{Results}
\label{sec:res}

A binned maximum likelihood fit is performed simultaneously to all SRs
and CRs using the CMS \textsc{Combine} package~\cite{CMS:2024onh}. The
fit uses inputs from the pNN output distributions in all regions
except the different-flavour ones, for which the dilepton \pt
distributions provide better discrimination between the $\PW\PW$ and
$\ttbar$ processes. The normalisations of the dominant backgrounds, including MisID events, are
constrained through the simultaneous fit to the CRs. Systematic
uncertainties are incorporated into the likelihood as nuisance
parameters. Uncertainties that only affect the normalisation of a
process are treated as log-normal nuisance parameters. Uncertainties
that affect the normalisation and shape of the distributions are
included through vertical morphing from different shape templates that
represent the $\pm$1 standard deviation variations, and are
incorporated into the likelihood with a Gaussian constraint term. The
full likelihood function is formed by taking the product of the
individual bin likelihoods and the probability density functions of
the nuisance parameters. The likelihood function, $\lumi$, that is
maximised is hence given by:
\begin{equation}
  \begin{split}
    \mathcal{L}(\mu, \boldsymbol{\theta}) = &  \prod_{i} \mathrm{P}(d_{i}^{\mathrm{SR}} | B_{i}^{\mathrm{SR}}(\boldsymbol{\theta}) + \mu S_{i}(\boldsymbol{\theta}) ) \\
    &\times \prod_{\mathrm{CR}} \Bigl( \prod_{i} \mathrm{P}(d_{i}^{\mathrm{CR}} | B_{i}^{\mathrm{CR}}(\boldsymbol{\theta})) \Bigr) \\
    &\times \prod_{j} \mathrm{P}(\theta_j),
  \end{split}
\end{equation}
where ${\mathrm{P}(x|y) = y^{x}\re^{-y}/x!}$ is the Poisson
distribution, and $d_i^X$ and $B_{i}^{\mathrm{X}}$ are the observed number of
events and the expected number of SM events, respectively, in bin $i$
of channel $X$. The index $i$ runs over all of the bins across each of two signal regions, and eight control regions, 
separately for the four data-taking years.
The signal term $S_i$ represents the expected number of
signal events in bin $i$, and $\mu$ denotes the signal strength
parameter. The symbol $\boldsymbol{\theta}$ represents the 
nuisance parameters that model systematic uncertainties that are
constrained by the Gaussian distributions denoted $\mathrm{P}(\theta_j)$. The expected
number of SM events, $B_{i}^{\mathrm{X}}$, can be decomposed into the separate
processes:
\begin{equation}
  \begin{split}
    B_{i}^{\mathrm{X}}  =&  \mu_{\PW\PW} N_{\PW\PW,\:i}^{\mathrm{X}}(\boldsymbol{\theta}) + \mu_{\ttbar} N_{\ttbar,\:i}^{\mathrm{X}}(\boldsymbol{\theta}) + \mu_{\PZ\PZ} N_{\PZ\PZ,\:i}^{\mathrm{X}}(\boldsymbol{\theta}) + \mu_{\PW\PZ} N_{\PW\PZ,\:i}^{\mathrm{X}}(\boldsymbol{\theta}) \\
    &  + \mu_{\mathrm{MisID}} N_{\mathrm{MisID},\:i}^{\mathrm{X}}(\boldsymbol{\theta}) + N_{\text{Other},\:i}^{\mathrm{X}}(\boldsymbol{\theta}),
  \end{split}
\end{equation}
where $N_{p,\:i}^{\mathrm{X}}$ is the expected number of events for process $p$,
$\mu_{\mathrm{p}}$ are the parameters that modify the total rate of each process, and $N_{\text{Other},\:i}^{\mathrm{X}}$ is the sum of the subdominant processes yields from MC, namely \DYjets, single top and triboson production.

\begin{figure}[ht!]
  \centering
  \includegraphics[width=0.49\textwidth]{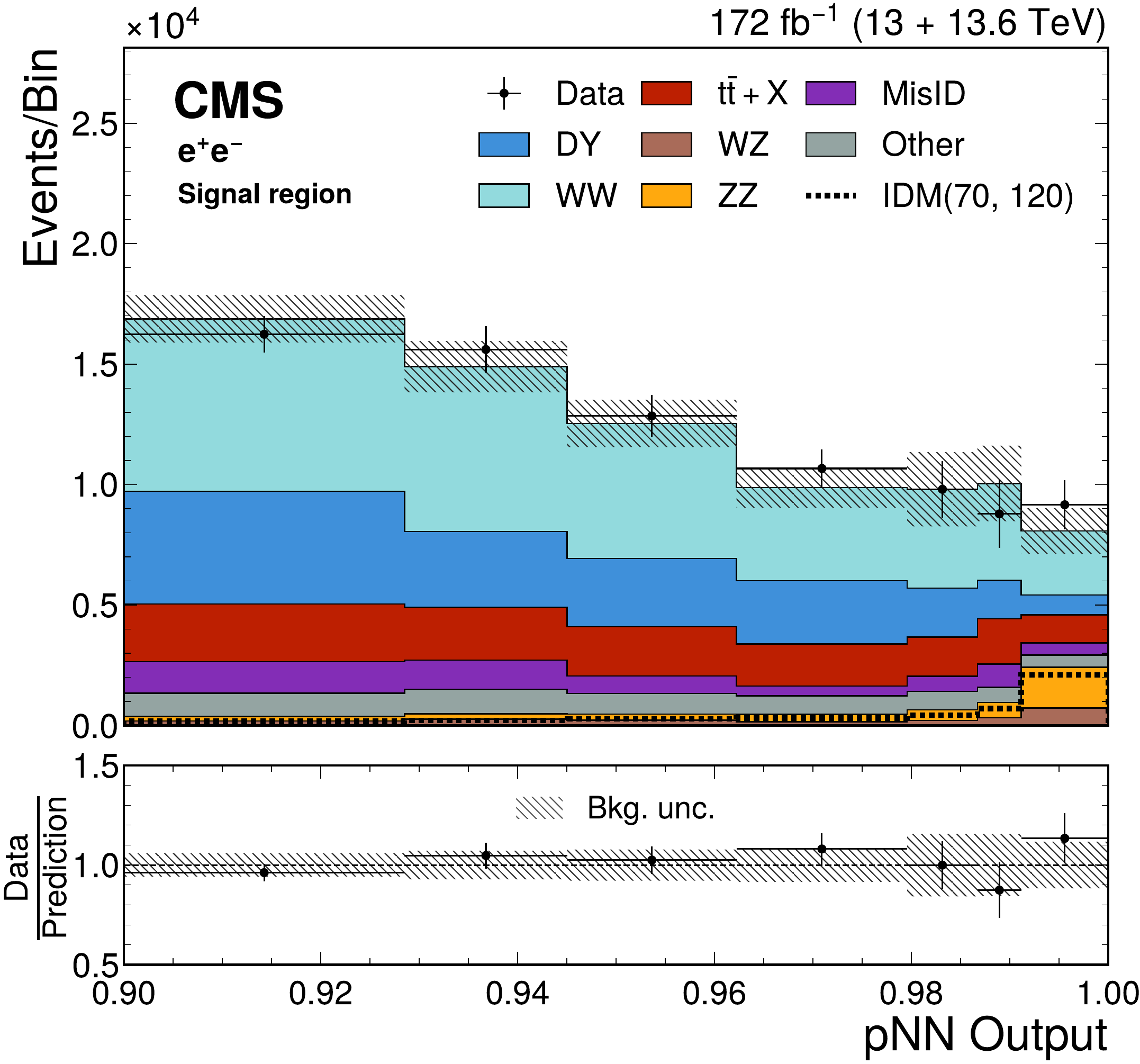}
  \includegraphics[width=0.49\textwidth]{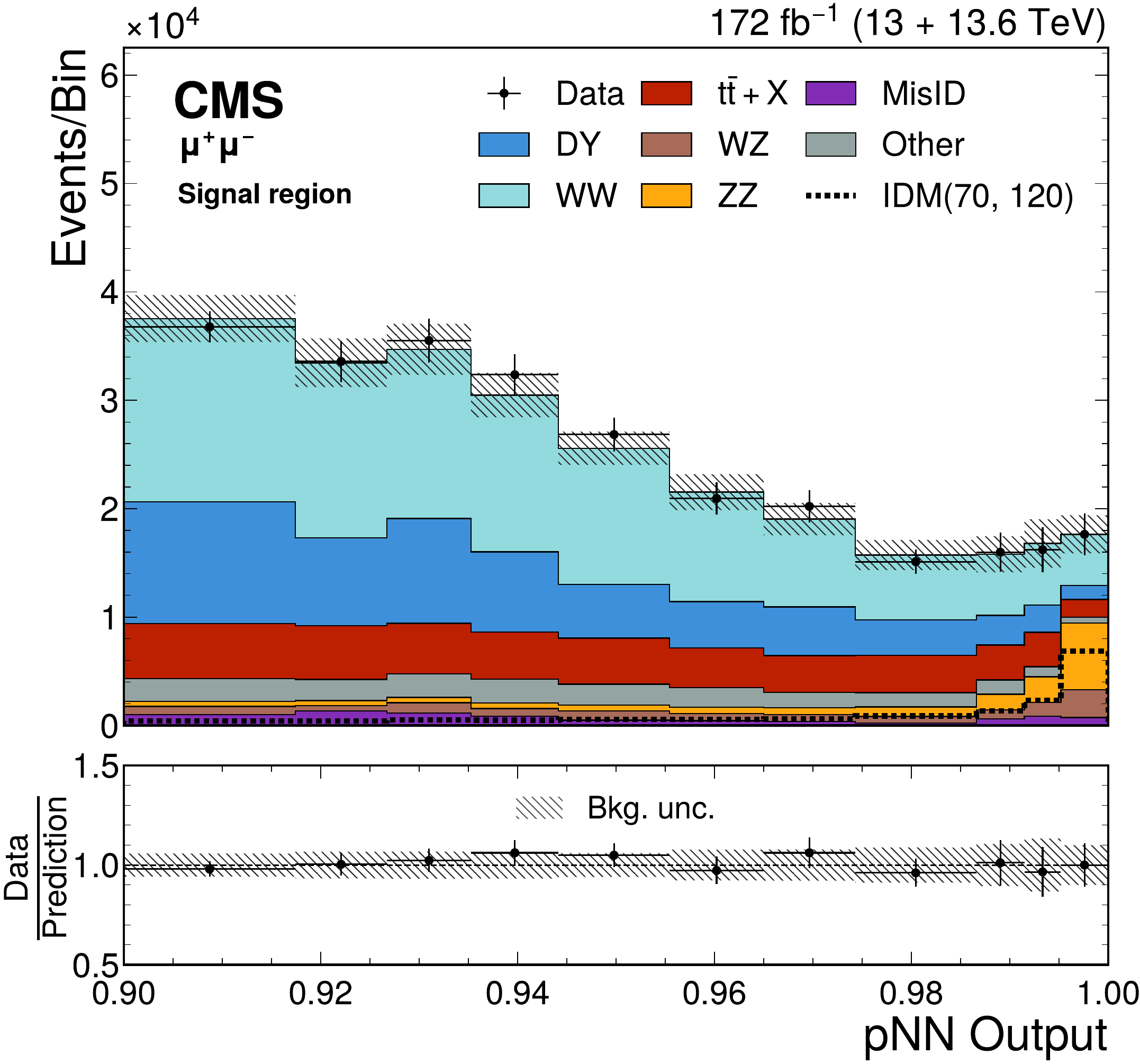}
  \caption{ Distributions of the pNN output for the data and the SM
    expectations in the SR after the background-only fit to the data
    in the (left) \ee channel and (right) \mumu channel for $\mh = 70$
    and $\ma = 120\GeV$. The dotted black line represents the signal,
    referred to as IDM$(\mh,\ma)$, with masses in units of \GeV. The
    lower panels show the ratio of data to the SM expectation. The
    error bars show data statistical uncertainties, while the hatched
    bands include the total uncertainty on the backgrounds.}
  \label{fig:post_fits_SR_mH70_mA120}
\end{figure}

\begin{figure}[ht!]
  \centering
  \includegraphics[width=0.49\textwidth]{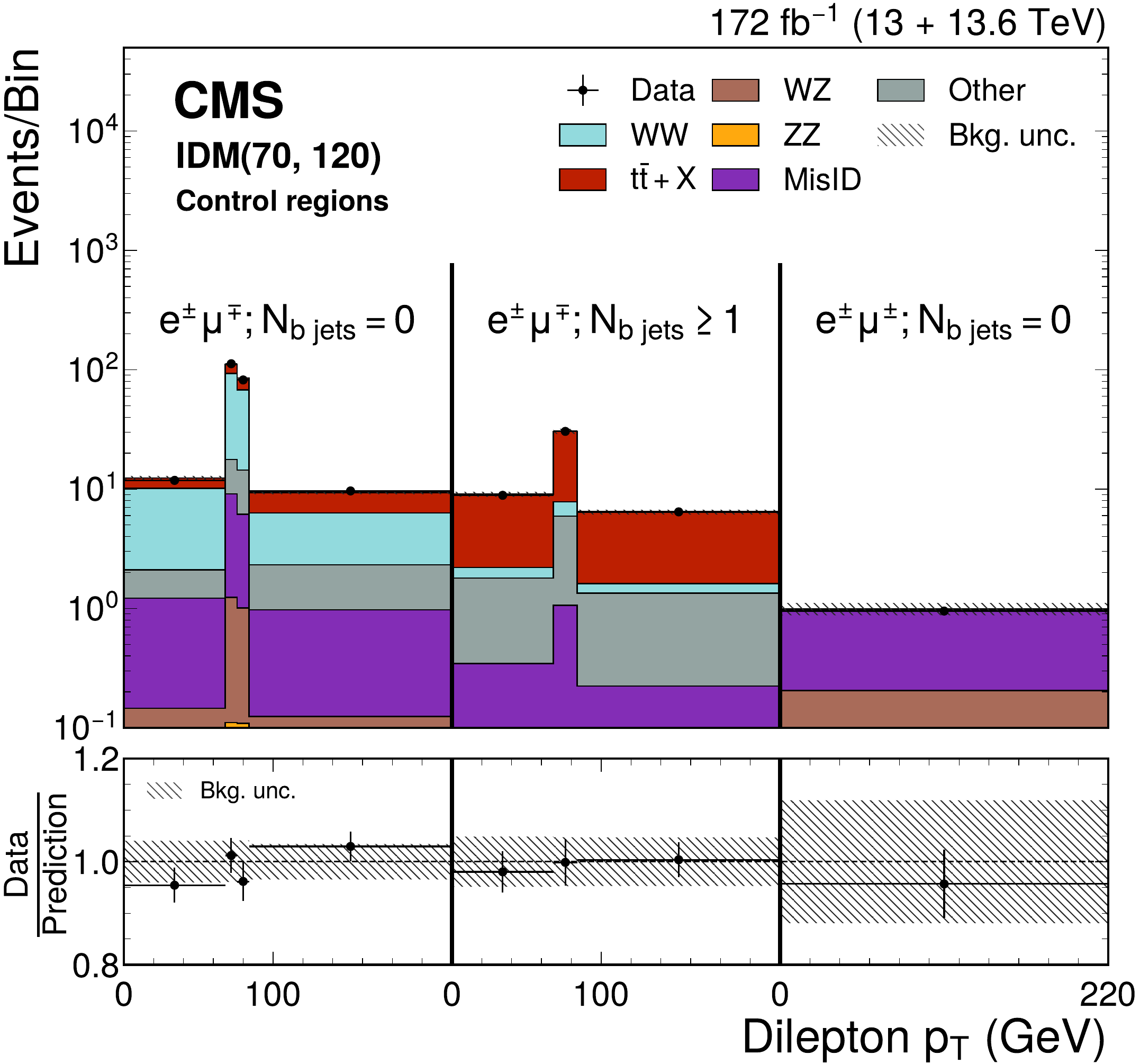}
  \includegraphics[width=0.49\textwidth]{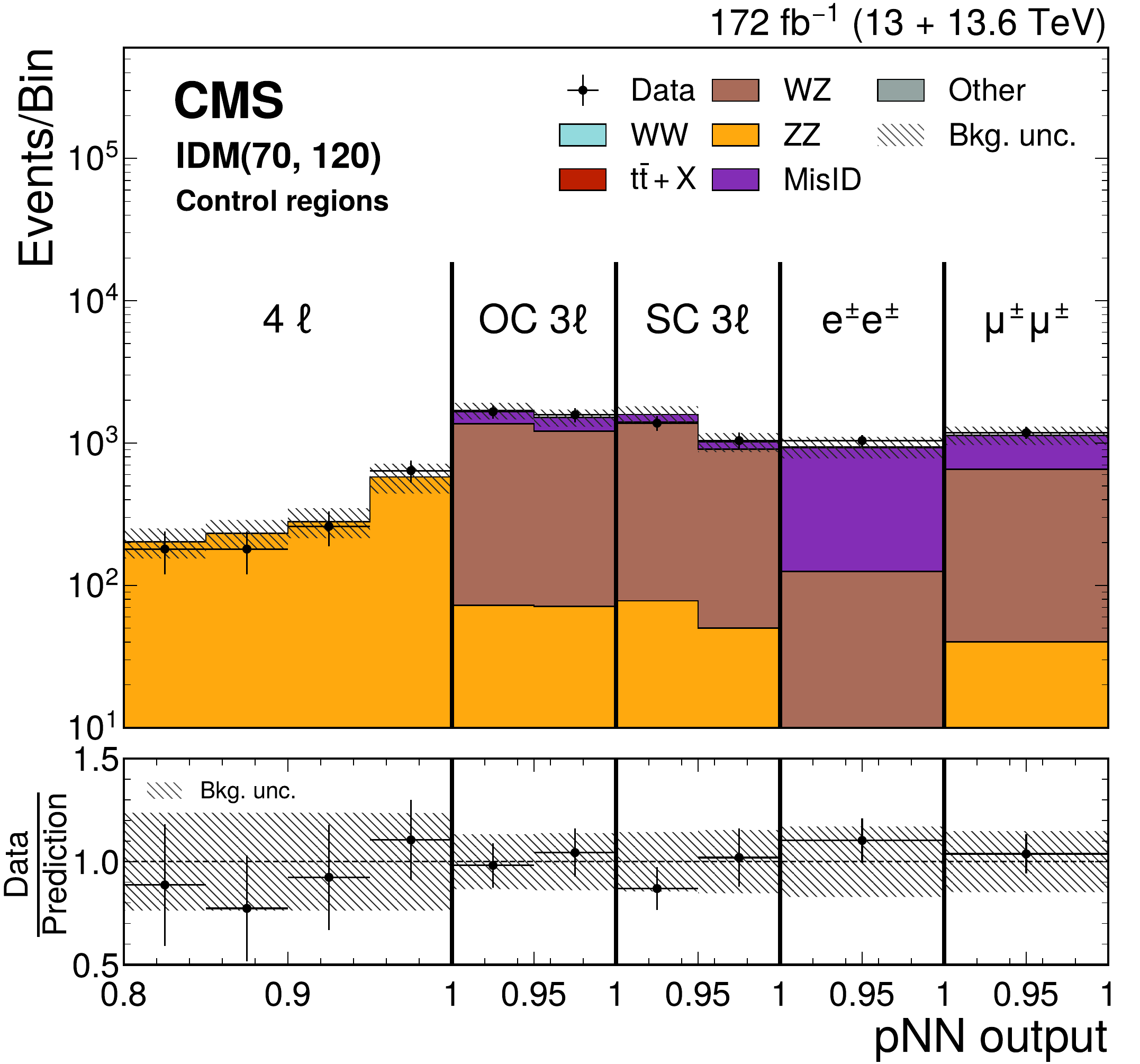}
  \caption{ Distributions of the data and the SM expectations in all
    CRs after the background-only fit to the data for $\mh = 70$ and
    $\ma = 120\GeV$. (Left) the dilepton \pt is used as the observable
    in the fit for the two $\PWm\PWp/\ttbar$ CRs as well as the
    different-flavour MisID CR, and (right) the pNN output is used for
    the $\PZ\PZ$ CR, both $\PW\PZ$ CRs (opposite-charge and
    same-charge) and the same-charge MisID CRs. The lower panels show
    the ratio of data to the SM expectation. The error bars show data
    statistical uncertainties, while the hatched bands include the
    total uncertainty on the backgrounds.}
  \label{fig:post_fits_mH70_mA120}
\end{figure}

\begin{figure}[ht!]
  \centering
  \includegraphics[width=0.49\textwidth]{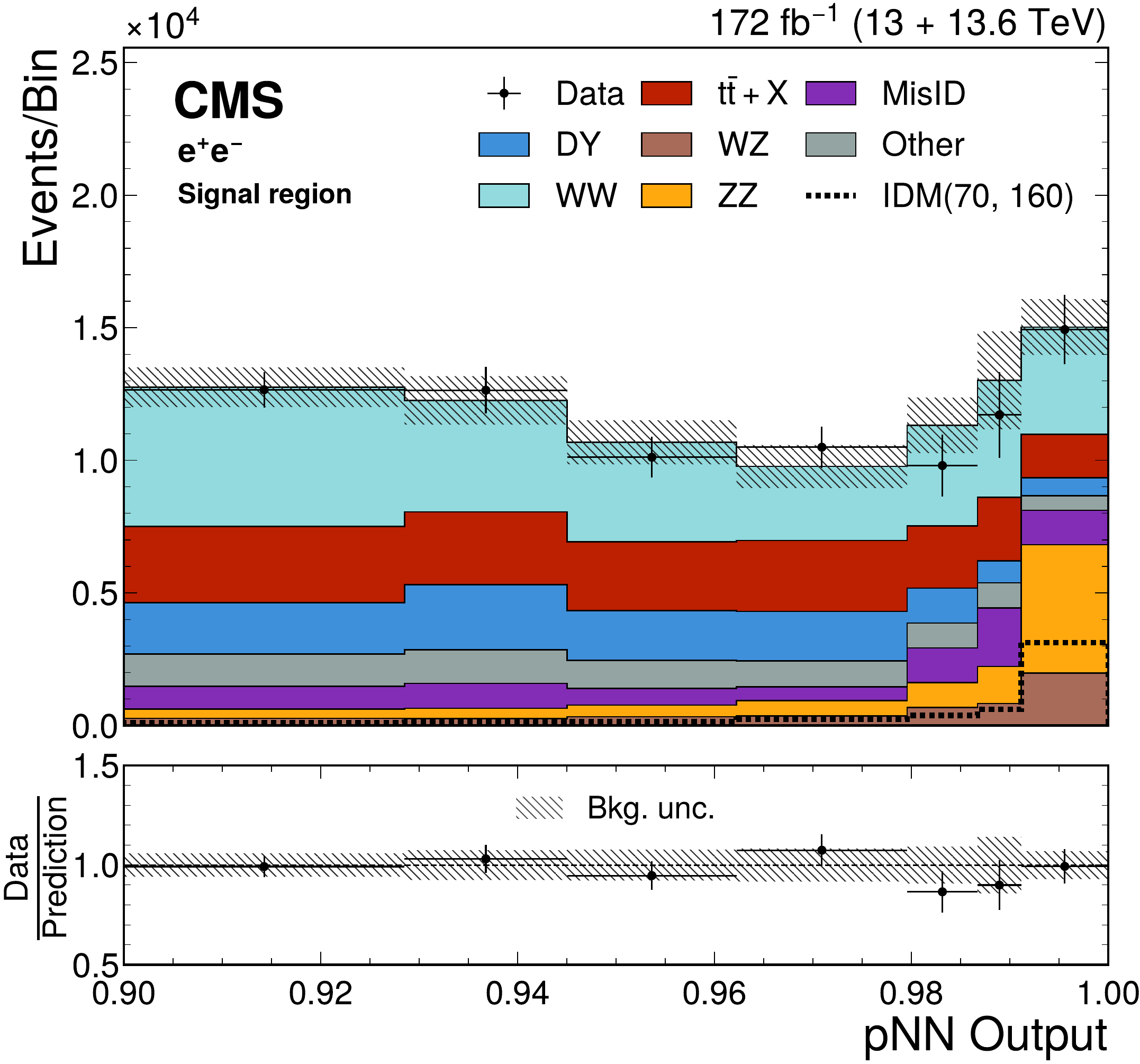}
  \includegraphics[width=0.49\textwidth]{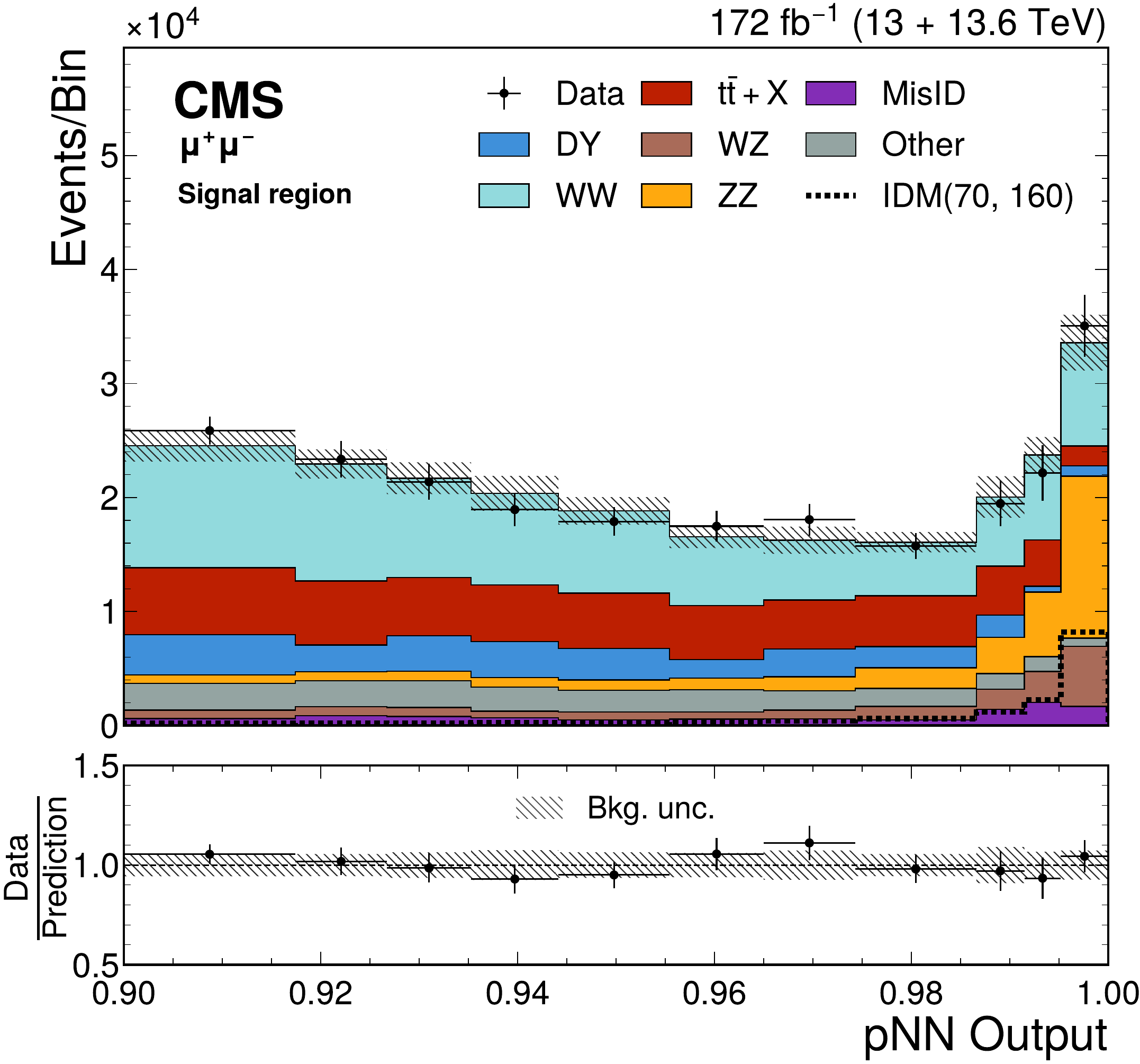}
  \caption{Distributions of the pNN output for the data and the SM
    expectations in the SR after the background-only fit to
    the data in the (left) \ee channel and (right) \mumu channel for
    $\mh = 70$ and $\ma = 160\GeV$. The dotted
    black line represents the signal, referred to as IDM$(\mh,\ma)$,
    with masses in units of \GeV. The lower panels show the ratio of
    data to the SM expectation. The error bars show data
    statistical uncertainties, while the hatched bands include the
    total uncertainty on the backgrounds.}
  \label{fig:post_fits_SR_mH70_mA160}
\end{figure}

\begin{figure}[ht!]
  \centering
  \includegraphics[width=0.49\textwidth]{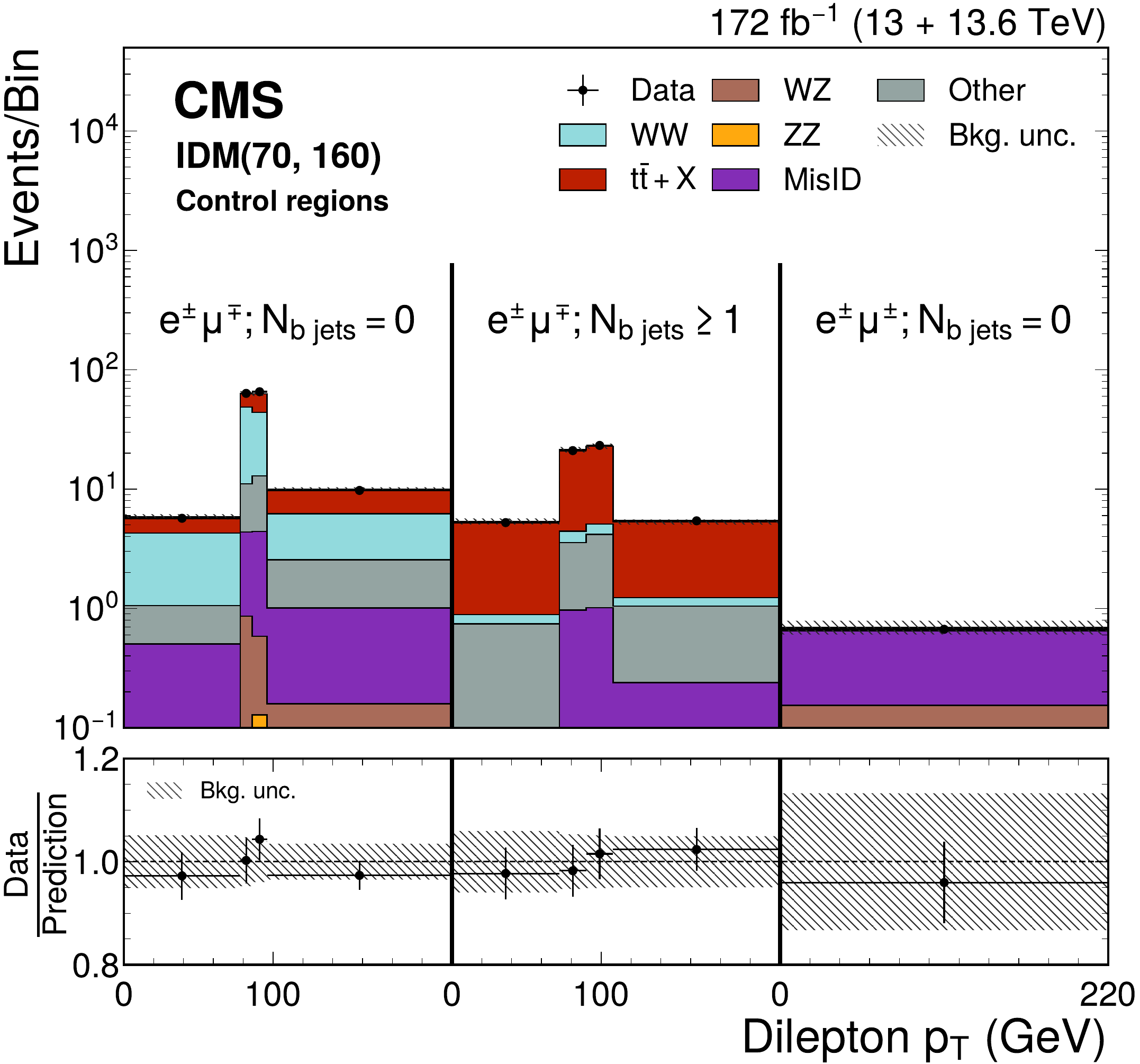}
  \includegraphics[width=0.49\textwidth]{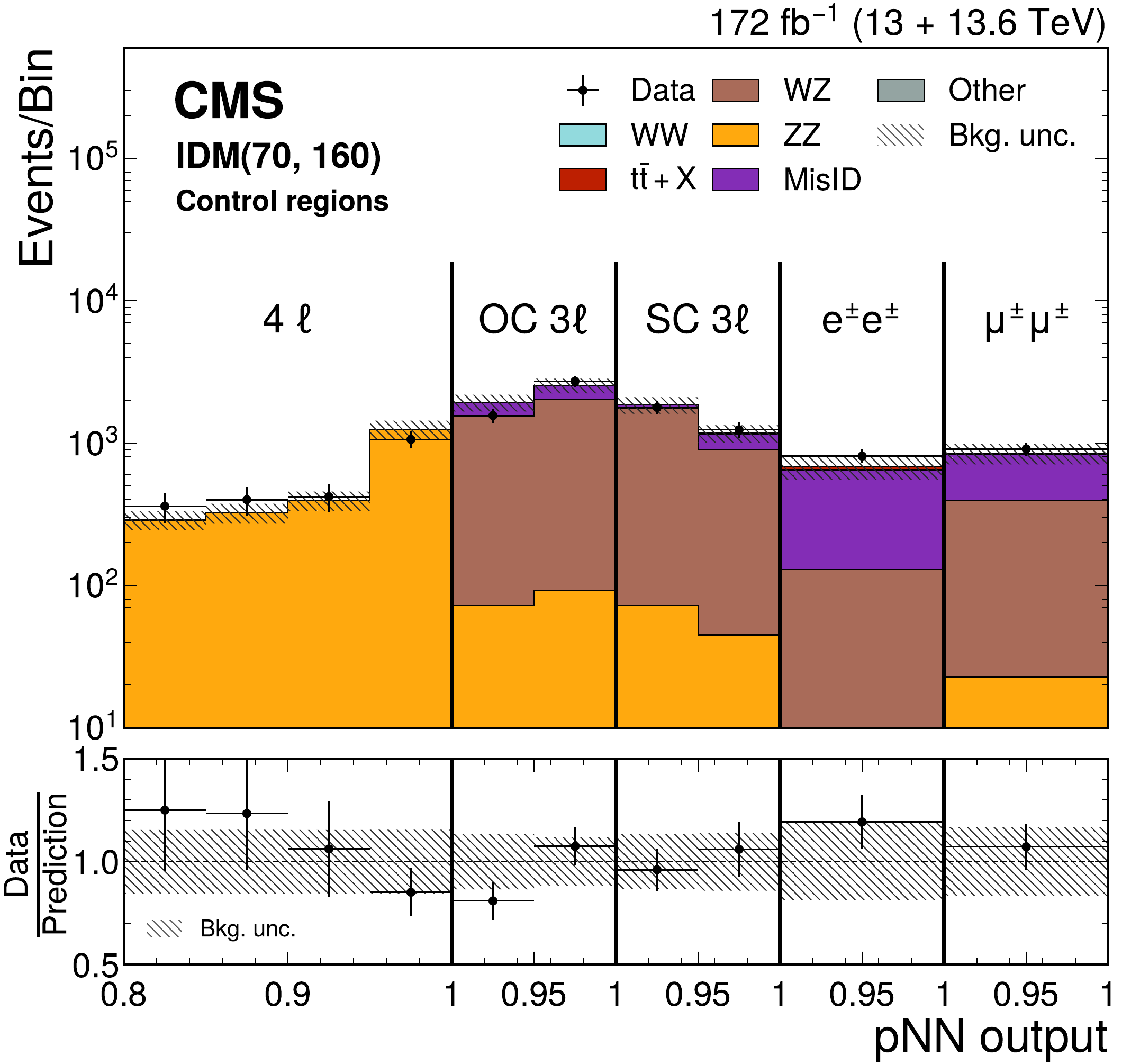}
  \caption{ Distributions of the data and the SM expectations in all
    CRs after the background-only fit to the data for $\mh = 70$ and
    $\ma = 160\GeV$. (Left) the dilepton \pt is used as the observable
    in the fit for the two $\PWm\PWp/\ttbar$ CRs as well as the
    different-flavour MisID CR, and (right) the pNN output is used for
    the $\PZ\PZ$ CR, both $\PW\PZ$ CRs (opposite-charge and
    same-charge) and the same-charge MisID CRs. The lower panels show
    the ratio of data to the SM expectation. The error bars show data
    statistical uncertainties, while the hatched bands include the
    total uncertainty on the backgrounds.}
  \label{fig:post_fits_mH70_mA160}
\end{figure}

A maximum likelihood fit to data is performed assuming a signal with $\mu=0$ and
allowing all other parameters to float. 
The distributions after this fit for mass points
$(\mh, \ma) = (70, 120)\GeV$ and $(70, 160)\GeV$ are shown in
Figs.~\ref{fig:post_fits_SR_mH70_mA120} to~\ref{fig:post_fits_mH70_mA160}.

Observed and expected 95\% confidence level (\CL) upper limits, and
local significances are calculated using profile likelihood ratio test
statistics~\cite{Cowan:2010js} with systematic uncertainties modelled
as nuisance parameters that are profiled, following the frequentist
approach. Asymptotic approximations of the test statistic
distributions are used~\cite{Cowan:2010js}.  No significant deviation
is observed between the SM background prediction and the data, with
the largest excess having a local p-value of approximately $1.5$
standard deviations, at $\mh=110\GeV$ and $\ma=145\GeV$.

\begin{figure}[ht!]
  \centering
  \includegraphics[width=0.65\textwidth]{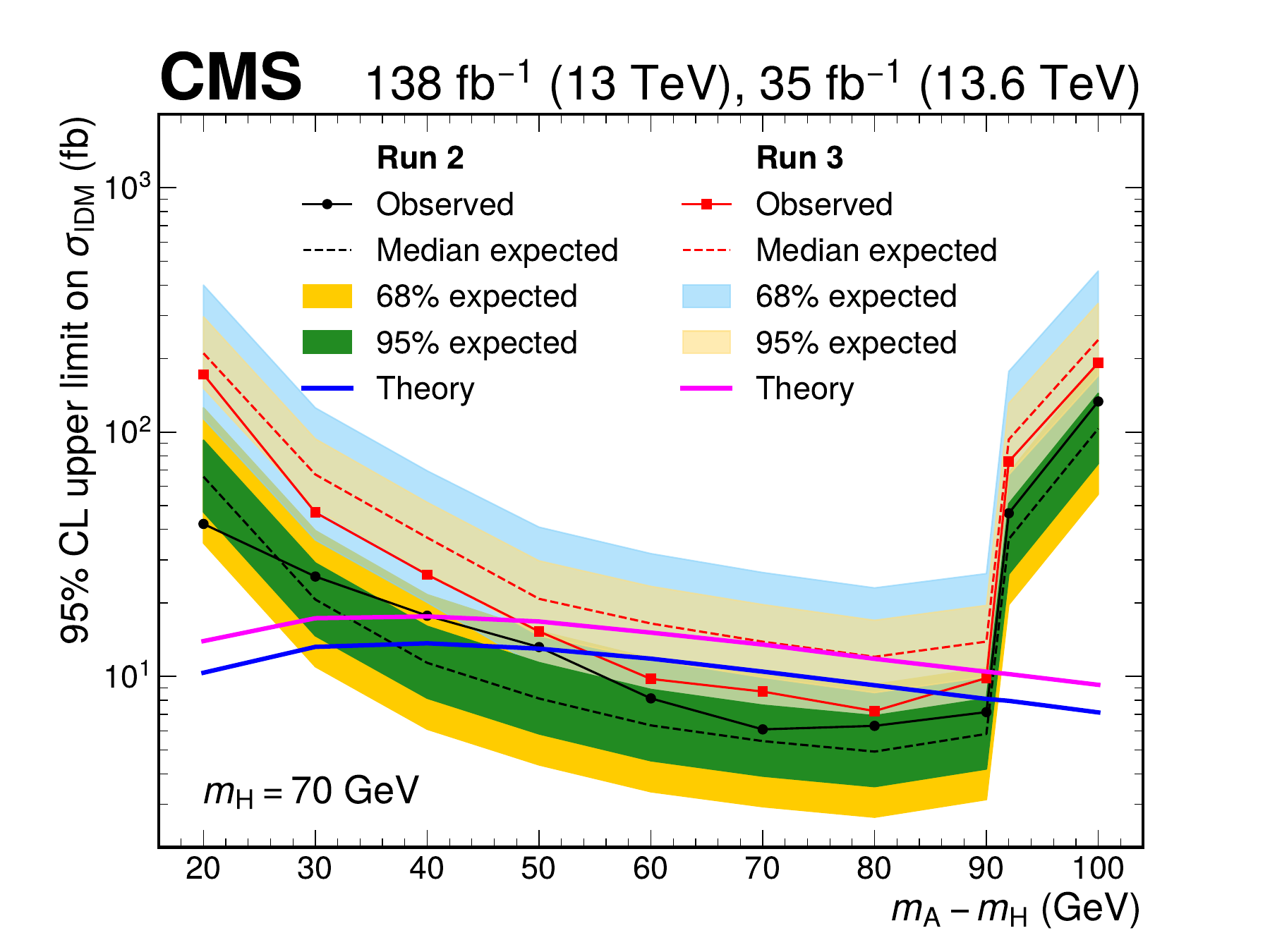}
  \caption{The 95\% \CL upper limit on $\sigma_{\mathrm{IDM}}$ as a
    function of $\ma - \mh$, for $\mh=70\GeV$, separately for the
    (black circles) Run 2 and (red squares) Run 3 data sets. Limits are
    calculated with the IDM parameters $\mch = \ma + 50\GeV$,
    $\lambda_2 = 1$, and $\lambda_{345} = 10^{-6}$. The limits,
    however, are insensitive to the choice of the $\lambda_2$ value,
    and to changes in $\mch$ and $\lambda_{345}$ within their allowed
    values. }
  \label{fig:1dlim}
\end{figure}

Upper limits are computed using the \CLs method detailed in
Refs.~\cite{Junk_1999,ALRead_2002}. The limits are calculated assuming
$\lambda_2 = 1$, $\lambda_{345} = 10^{-6}$ and $\mch = \ma + 50\GeV$;
however the results are insensitive to changes in these parameters
within their allowed ranges. The 95\% \CL observed and expected upper
limits on the sum of the signal cross sections for the production of
$\PH\PH\Pell\Pell$ and $\PH\PH\Pell\Pell\PGn\PAGn$,
$\sigma_{\mathrm{IDM}}$, are shown as a function of $\mdiff$ for $\mh
= 70\GeV$ in Fig.~\ref{fig:1dlim}, separately for the Run 2 and Run 3
data sets. The sudden reduction in sensitivity between 90\GeV and
92\GeV is due to the $\PZ$ boson being produced on-shell. Different
choices for the value of \mh have little impact on these limits;
tabulated values over the parameter space considered in this paper can
be found in Ref.~\cite{hepdata}. However the signal cross section
values vary significantly with \mh (see
Fig.~\ref{fig:parameter_scan_mH_mA-mH}). The data sets are then
combined to extract exclusion contours in the (\mh, \ma) plane. While
the signal cross section at \sqs{13.6} is larger than that at
\sqs{13}, the amount of data used from Run 3 is much smaller than that
from Run 2 meaning the sensitivity is dominated by the latter. However, as the Run 2 data present an excess and the Run 3 data a deficit, both within the 68\% quantiles, the observed exclusions become more similar.
Table~\ref{table:systimpact} summarises the uncertainties in the
extracted signal strength, separating the contributions from different
groups of uncertainties. The theory group contains uncertainties on
the modelling of the \Zjets contribution, due to PDFs, renormalisation
and factorisation scales. The leading uncertainty is the statistical uncertainty 
of the data, dominated by the $\PZ\PZ$ CR. Among the systematic sources, the 
simulated event counts and those related to jet calibration are the largest.

\begin{table}[ht!]
    \centering \topcaption{Uncertainty breakdown in the fitted signal
      strength for two signal mass points. The sources of uncertainty
      are separated into different groups.}
    \renewcommand{\arraystretch}{1.2}
    \begin{tabular}{l  c  c }
       & \multicolumn{2}{c}{$(\mh, \ma)$\GeV}\\
      Group of systematic uncertainties & $(70, 120)$ & $(70, 160)$\\
      \hline
    Theory &  $\pm 0.04$ & $\pm0.04$\\
    Simulated event count &  $\pm 0.09$ &  $\pm 0.08$\\
    Jet calibration &  $+0.07,-0.06$ & $\pm 0.04$ \\
    Lepton, jet ID and trigger efficiency  &  $+0.03,-0.04$ & $\pm 0.04$\\
    Integrated luminosity/pileup   &  $\pm 0.04$ & $\pm 0.02$ \\
    Statistical uncertainty &  $+0.26,-0.25$ & $+0.31,-0.30$ \\
    \hline
    Total uncertainty & $+0.29,-0.28$ & $\pm 0.32$ \\
    \end{tabular}
    \label{table:systimpact}
\end{table}

\begin{figure}[ht!]
  \centering
  \includegraphics[width=0.65\textwidth]{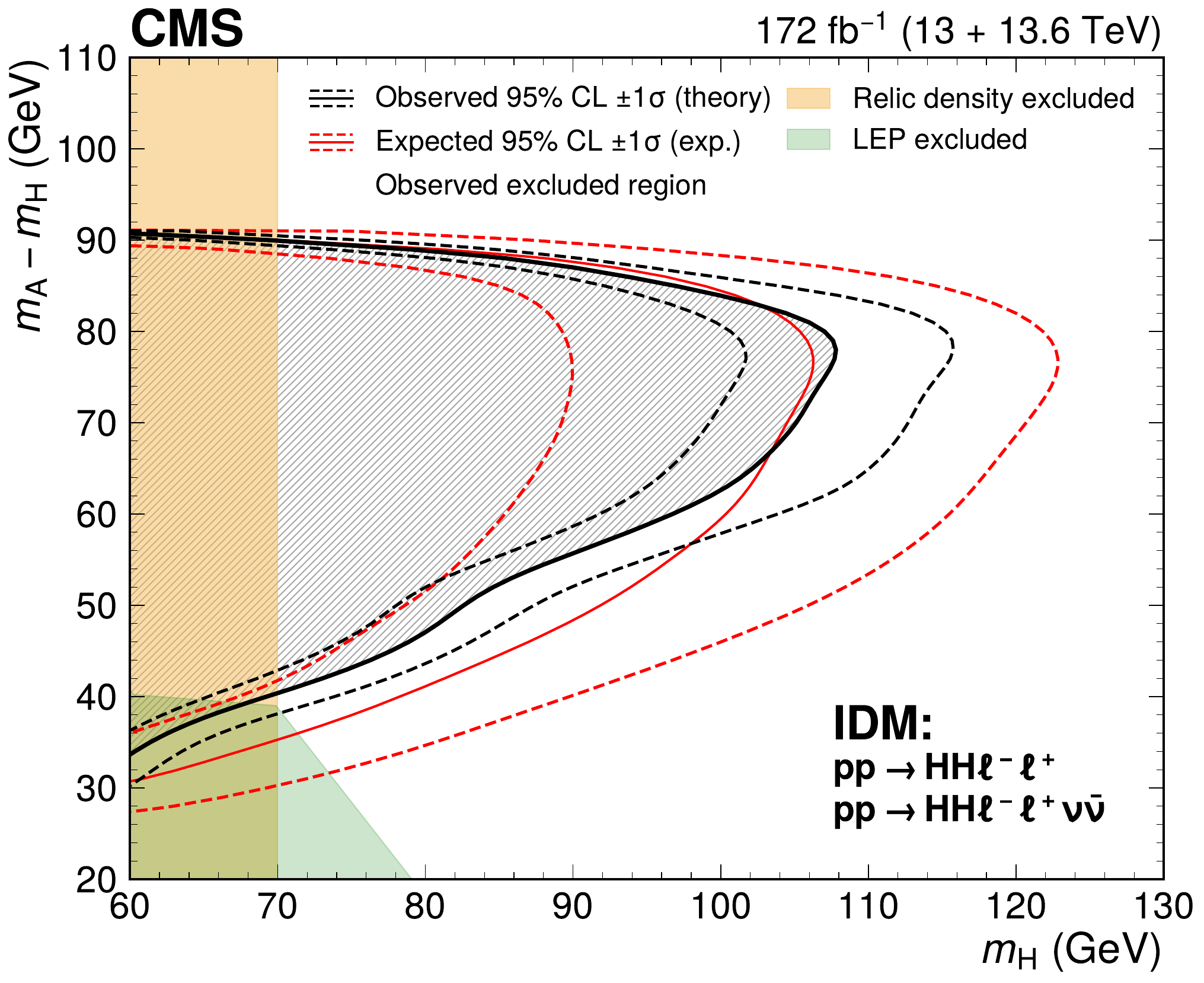}
  \caption{The 95\% \CL exclusion limits in terms of $\mh$ and $\ma - \mh$.  The red
    dashed lines indicate the $\pm1$ standard deviation bands from experimental
    uncertainties, whilst the black dashed lines indicate the
    $\pm1$ standard deviation bands from theoretical uncertainties in the signal
    samples. The exclusion limits from the LEP reinterpretation and
    relic density constraints are overlaid in green and yellow,
    respectively (see Section~\ref{sec:idm}). Limits are calculated with
    the IDM parameters $\mch = \ma + 50\GeV$, $\lambda_2 = 1$, and
    $\lambda_{345} = 10^{-6}$. The limits, however, are insensitive to
    the choice of the $\lambda_2$ value, and to
    changes in $\mch$ and $\lambda_{345}$ within their allowed values. }
  \label{fig:limits}
\end{figure}

The 95\% \CL expected exclusion limits on $\mh$ and $\ma$ are shown in
Fig.~\ref{fig:limits}, in terms of $\mh$ and $\ma-\mh$. The observed
limits are consistent with the expected limits, lying within the 68\%
quantiles shown by the red dashed lines. The black dashed lines show
the change in the observed limit due to variation of the signal cross
sections within their theoretical uncertainties. The hatched area
indicates the excluded region from the observed 95\% \CL limits, which
extends up to a maximum DM mass of $\mh=108\GeV$ at a mass splitting
of $\ma - \mh = 78\GeV$. The observed (expected) exclusion region
extends up to $\mdiff=90\,(90)\GeV$ and down to $\mdiff=40\,(35)\GeV$
at $\mh=70\GeV$. Overlaid are the current limits based on a LEP
reinterpretation and relic density bounds. These exclusion limits
significantly extend the constraints from previous direct and indirect
measurements, and DM searches.

\section{Summary}
\label{sec:concl}

The Inert Doublet Model predicts additional scalars, including two
neutral scalars $\PH$ and $\PA$, which couple only to bosons. The
lightest neutral scalar, $\PH$, is stable and provides a viable dark
matter candidate. The pair production of such new scalars is
investigated in a final state containing two electrons or two
muons. The search is performed using proton-proton collisions at
\sqs{13} and $13.6\TeV$, corresponding to integrated luminosities of
\invlumiII\fbinv and \invlumiIII\fbinv, delivered by the LHC and
recorded by the CMS experiment between 2016 and 2018, and in 2022,
respectively. After a preselection to remove the largest standard
model backgrounds, a parameterised neural network is trained to
discriminate the different signal mass points from the remaining
backgrounds. Dedicated control regions for each of the dominant
backgrounds are constructed. A simultaneous fit of the signal region
together with the control regions is used to set 95\% confidence level
exclusion limits on the signal production cross section in the \mh vs.
\mdiff plane. The observed (expected) exclusion region reaches
$\mh=108\,(106)\GeV$ for $\mdiff=78\,(76)\GeV$ and, at $\mh=70\GeV$,
covers the range of $\mdiff=40$--90\,(35--90)\GeV. These results
represent the first limits on the masses of the neutral scalars in the
Inert Doublet Model obtained by a dedicated search using collision
data. These exclusion limits significantly extend the constraints from
previous direct and indirect measurements and dark matter searches.

 \begin{acknowledgments}
 We congratulate our colleagues in the CERN accelerator departments for the excellent performance of the LHC and thank the technical and administrative staffs at CERN and at other CMS institutes for their contributions to the success of the CMS effort. In addition, we gratefully acknowledge the computing centres and personnel of the Worldwide LHC Computing Grid and other centres for delivering so effectively the computing infrastructure essential to our analyses. Finally, we acknowledge the enduring support for the construction and operation of the LHC, the CMS detector, and the supporting computing infrastructure provided by the following funding agencies: SC (Armenia), BMBWF and FWF (Austria); FNRS and FWO (Belgium); CNPq, CAPES, FAPERJ, FAPERGS, and FAPESP (Brazil); MES and BNSF (Bulgaria); CERN; CAS, MoST, and NSFC (China); MINCIENCIAS (Colombia); MSES and CSF (Croatia); RIF (Cyprus); SENESCYT (Ecuador); ERC PRG and PSG, TARISTU24-TK10 and MoER TK202 (Estonia); Academy of Finland, MEC, and HIP (Finland); CEA and CNRS/IN2P3 (France); SRNSF (Georgia); BMFTR, DFG, and HGF (Germany); GSRI (Greece); MATE and NKFIH (Hungary); DAE and DST (India); IPM (Iran); SFI (Ireland); INFN (Italy); MSIT and NRF (Republic of Korea); MES (Latvia); LMTLT (Lithuania); MOE and UM (Malaysia); BUAP, CINVESTAV, CONACYT, LNS, SEP, and UASLP-FAI (Mexico); MOS (Montenegro); MBIE (New Zealand); PAEC (Pakistan); MSHE, NSC, and NAWA (Poland); FCT (Portugal); MESTD (Serbia); MICIU/AEI and PCTI (Spain); MOSTR (Sri Lanka); Swiss Funding Agencies (Switzerland); MST (Taipei); MHESI (Thailand); TUBITAK and TENMAK (T\"{u}rkiye); NASU (Ukraine); STFC (United Kingdom); DOE and NSF (USA).

\begin{sloppypar}
\setlength\emergencystretch{\hsize}
\hyphenation{Rachada-pisek} Individuals have received support from the Marie-Curie programme and the European Research Council and Horizon 2020 Grant, contract Nos.\ 675440, 724704, 752730, 758316, 765710, 824093, 101115353, 101002207, 101001205, and COST Action CA16108 (European Union); the Leventis Foundation; the Alfred P.\ Sloan Foundation; the Alexander von Humboldt Foundation; the Science Committee, project no. 22rl-037 (Armenia); the Fonds pour la Formation \`a la Recherche dans l'Industrie et dans l'Agriculture (FRIA) and Fonds voor Wetenschappelijk Onderzoek contract No. 1228724N (Belgium); the Beijing Municipal Science \& Technology Commission, No. Z191100007219010, the Fundamental Research Funds for the Central Universities, the Ministry of Science and Technology of China under Grant No. 2023YFA1605804, the Natural Science Foundation of China under Grant No. 12535004, and USTC Research Funds of the Double First-Class Initiative No.\ YD2030002017 (China); the Ministry of Education, Youth and Sports (MEYS) of the Czech Republic; the Shota Rustaveli National Science Foundation (Georgia); the Deutsche Forschungsgemeinschaft (DFG), among others, under Germany's Excellence Strategy -- EXC 2121 ``Quantum Universe" -- 390833306, and under project number 400140256 - GRK2497; the Hellenic Foundation for Research and Innovation (HFRI), Project Number 2288 (Greece); the Hungarian Academy of Sciences, the New National Excellence Program - \'UNKP, the NKFIH research grants K 131991, K 138136, K 143460, K 143477, K 147557, K 146913, K 146914, K 147048, TKP2021-NKTA-64, and 2025-1.1.5-NEMZ\_KI-2025-00004, and MATE KKP and KKPCs Research Excellence and Flagship Research Groups grants (Hungary); the Council of Science and Industrial Research, India; ICSC -- National Research Centre for High Performance Computing, Big Data and Quantum Computing, FAIR -- Future Artificial Intelligence Research, and CUP I53D23001070006 (Mission 4 Component 1), funded by the NextGenerationEU program, the Italian Ministry of University and Research (MUR) under Bando PRIN 2022 -- CUP I53C24002390006, PRIN PRIMULA 2022RBYK7T (Italy); the Latvian Council of Science; the Ministry of Science and Higher Education, project no. 2022/WK/14, and the National Science Centre, contracts Opus 2021/41/B/ST2/01369, 2021/43/B/ST2/01552, 2023/49/B/ST2/03273, and the NAWA contract BPN/PPO/2021/1/00011 (Poland); the Funda\c{c}\~ao para a Ci\^encia e a Tecnologia (Portugal); the National Priorities Research Program by Qatar National Research Fund; MICIU/AEI/10.13039/501100011033, ERDF/EU, ``European Union NextGenerationEU/PRTR", projects PID2022-142604OB-C21, PID2022-139519OB-C21, PID2023-147706NB-I00, PID2023-148896NB-I00, PID2023-146983NB-I00, PID2023-147115NB-I00, PID2023-148418NB-C41, PID2023-148418NB-C42, PID2023-148418NB-C43, PID2023-148418NB-C44, PID2024-158190NB-C22, RYC2021-033305-I, RYC2024-048719-I, CNS2023-144781, CNS2024-154769 and Plan de Ciencia, Tecnolog{\'i}a e Innovaci{\'o}n de Asturias, Spain; the Chulalongkorn Academic into Its 2nd Century Project Advancement Project, the National Science, Research and Innovation Fund program IND\_FF\_68\_369\_2300\_097, and the Program Management Unit for Human Resources \& Institutional Development, Research and Innovation, grant B39G680009 (Thailand); the Eric \& Wendy Schmidt Fund for Strategic Innovation through the CERN Next Generation Triggers project under grant agreement number SIF-2023-004; the Kavli Foundation; the Nvidia Corporation; the SuperMicro Corporation; the Welch Foundation, contract C-1845; and the Weston Havens Foundation (USA).
\end{sloppypar}
\end{acknowledgments}\section*{Data availability} Release and preservation of data used by the CMS Collaboration as the basis for publications is guided by the  \href{https://doi.org/10.7483/OPENDATA.CMS.1BNU.8V1W}{CMS data preservation, re-use and open access policy}.

\bibliography{auto_generated}

\cleardoublepage \appendix\section{The CMS Collaboration \label{app:collab}}\begin{sloppypar}\hyphenpenalty=5000\widowpenalty=500\clubpenalty=5000\input{NPS-25-002-public-authorlist.tex}\end{sloppypar}
%%% END EDITABLE REGION %%%
% skeleton_end
\end{document}

%% file: NPS-25-002-public-authorlist.tex
\cmsinstitute{Yerevan Physics Institute, Yerevan, Armenia}
{\tolerance=6000
A.~Gevorgyan\cmsorcid{0000-0003-2751-9489}, A.~Hayrapetyan, V.~Makarenko\cmsorcid{0000-0002-8406-8605}, A.~Tumasyan\cmsAuthorMark{1}\cmsorcid{0009-0000-0684-6742}
\par}
\cmsinstitute{Marietta Blau Institute for Particle Physics, Vienna, Austria}
{\tolerance=6000
W.~Adam\cmsorcid{0000-0001-9099-4341}, L.~Benato\cmsorcid{0000-0001-5135-7489}, T.~Bergauer\cmsorcid{0000-0002-5786-0293}, M.~Dragicevic\cmsorcid{0000-0003-1967-6783}, S.~Gundacker\cmsorcid{0000-0003-2087-3266}, P.S.~Hussain\cmsorcid{0000-0002-4825-5278}, M.~Jeitler\cmsAuthorMark{2}\cmsorcid{0000-0002-5141-9560}, N.~Krammer\cmsorcid{0000-0002-0548-0985}, A.~Li\cmsorcid{0000-0002-4547-116X}, D.~Liko\cmsorcid{0000-0002-3380-473X}, M.~Matthewman, J.~Schieck\cmsAuthorMark{2}\cmsorcid{0000-0002-1058-8093}, R.~Sch\"{o}fbeck\cmsAuthorMark{2}\cmsorcid{0000-0002-2332-8784}, M.~Shooshtari\cmsorcid{0009-0004-8882-4887}, M.~Sonawane\cmsorcid{0000-0003-0510-7010}, N.~Van~Den~Bossche\cmsorcid{0000-0003-2973-4991}, W.~Waltenberger\cmsorcid{0000-0002-6215-7228}, C.-E.~Wulz\cmsAuthorMark{2}\cmsorcid{0000-0001-9226-5812}
% ...
\par}
\vskip\cmsinstskip
\dag:~Deceased\\
$^{1}$Also at Yerevan State University, Yerevan, Armenia\\
$^{2}$Also at TU Wien, Vienna, Austria\\
% ...
